\documentclass[10pt,draft,reqno]{amsart}
     \makeatletter
     \def\section{\@startsection{section}{1}%
     \z@{.7\linespacing\@plus\linespacing}{.5\linespacing}%
     {\bfseries
     \centering
     }}
     \def\@secnumfont{\bfseries}
     \makeatother
\newtheorem{theorem}{Theorem}[section]
\newtheorem{lemma}[theorem]{Lemma}
\newtheorem{proposition}[theorem]{Proposition}

\theoremstyle{definition}

\newtheorem{example}[theorem]{Example}
\theoremstyle{remark}
\newtheorem{remark}[theorem]{Remark}

\numberwithin{equation}{section}
\newcommand{\real}{\mathbb{R}}
\newcommand{\sig}{\mathcal{F}}
\begin{document}

\title[An Option Pricing Model with Memory]{An Option Pricing Model with Memory}

\author{Flavia Sancier}
\address{Flavia Sancier: Sciences Division, Antioch College, Yellow Springs, OH 45387, USA}
\email{fsancier@antiochcollege.edu}
\urladdr{https://sites.google.com/a/antiochcollege.edu/fsancier/}

\author[Salah Mohammed]{Salah Mohammed}
\address{Salah Mohammed: Deceased December 21, 2016}

\subjclass[2010] {
	34K50, 
	60H30, 
	91G20.} 

\keywords{Option pricing, stochastic functional differential equations, Black-Scholes, stochastic volatility.}

\begin{abstract}

We obtain option pricing formulas for  stock price models in which the drift and volatility terms are functionals of a continuous history of the stock prices.
That is, the stock dynamics follows a nonlinear stochastic functional differential equation. A model with full memory is obtained  via approximation through a stock price model in which the continuous  path dependence does not go up to the present: there is a \emph{memory gap}. A strong solution is obtained by \emph{closing the gap}. Fair option prices are obtained through an equivalent (local) martingale measure via Girsanov's Theorem and therefore are given in terms of a conditional expectation. The models maintain the completeness of the market and have no arbitrage opportunities. 
\end{abstract}

\maketitle

\section{Introduction}

The development of a theory for pricing options in financial markets has its roots in the early 1900's with Bachelier \cite{bachelier}, who initiated and used the theory of Brownian motion for modeling stock prices. 
But it was not before the 1960's that major results in mathematical finance were obtained by Samuelson \cite{samuelson1,samuelson2}, who used geometric Brownian motion to model the random behavior of stock prices and also developed the idea that discounted prices follow a martingale. 

In 1973, the well-known  \emph{Black-Scholes} model \cite{BS} was presented, together with Merton's \emph{Theory of rational option pricing} \cite{merton}. 
The main assumptions of the Black-Scholes model are that the stock price follows a geometric Brownian motion with \emph{constant} volatility and that there are no \emph{arbitrage opportunities}.

Despite Black and Scholes' extraordinary achievement, tests of their model on real market data have questioned the assumption of constant volatility in the stock dynamics (e.g, Scott \cite{scott}, Johnson and Shanno \cite{johnson}). Indeed, the presence of \emph{smiles} in the graph of \emph{implied volatility} versus \emph{strike price} (Bates \cite{bates}) suggests that the idea of constant volatility does not fit real data. For this reason, several variants of the Black-Scholes model with non-constant volatility have been proposed (e.g., Cox and Ross \cite{cox}, Hobson and Rogers\cite{Hobson}). 

In the present work, we take into account the possible dependence of the stock dynamics on its history. 
This is a reasonable consideration since decision makers take into account their knowledge of the past market behavior when selling or  purchasing assets.

In option pricing theory, several authors have proposed models with hereditary structure (e.g., Hobson and Rogers \cite{Hobson}, Arriojas, Hu, Mohammed and Pap \cite{bs1}, Stoica \cite{stoica}, Kazmerchuk \cite{kaz}, Chang \cite{chang}, Lee \cite{lee}). 

We derive option pricing formulas for two stock dynamics described by nonlinear stochastic functional differential equations. First we introduce a stock price model with a \emph{memory gap} as an extension of \cite{bs1}. Solutions of systems with a memory gap are processes in which the continuous dependence of the state on its history goes only up to a specific time in the past. In this way, there is a \emph{gap} between the past and present states. Although more restrictive in its past dependence, this stock dynamics has more  relaxed conditions on the drift and volatility terms, viz. no Lipschitz condition is needed for existence and uniqueness of strong solutions. 

The second stock price model has full finite memory and its drift and volatility terms are uniformly bounded and globally Lipschitz. It is similar to the stock dynamics introduced in \cite{chang}. 
We show that strong solutions of the stock price model with memory gap converge to solutions of the model with full finite memory as the gap goes to zero. Option pricing formulas are obtained for both models using such convergence. 

Since the option pricing formulas are derived through an equivalent (local) martingale measure via Girsanov's Theorem, they take the form of a conditional expectation, which makes them computationally simple to simulate through the use of Monte Carlo methods. 

The paper is outlined as follows. In section \ref{dp}, we introduce the stock price model with memory gap and show its existence and uniqueness. 
In section \ref{ndp}, we show that the model with memory gap converges to the model with full finite memory as the gap goes to zero. In section \ref{opdm}, we derive an option pricing formula for the stock dynamics with memory gap and finally, in section \ref{opfm}, we derive an option pricing formula for the stock price model with full finite memory. 
Its derivation is based on an equivalent (local) martingale measure via Girsanov's Theorem \cite{girsanov}. Therefore, the formula is given in terms of a conditional expectation. The model maintains the completeness of the market, has no arbitrage opportunities and its volatility has intrinsic randomness.

\section{A stock price model with memory gap}\label{dp}

In this section we present a stock price model in which the drift and volatility terms depend on a finite history of the stock prices up to a specific time in the past.
The model is an extension of \cite{bs1}. 

\subsection{Framework}\label{framework2}
Let $(\Omega,\sig,(\sig_t)_{t\in[0,T]},P)$ be a filtered probability space satisfying the usual conditions. Denote by $C:=C([-L,0],\real)$ the Banach space of all continuous paths $\eta:[-L,0]\rightarrow \real$ given the supremum norm. Let $f:[0,T]\times C\rightarrow \real$ and $g:[0,T]\times C\rightarrow \real$ be jointly continuous  functionals, and consider an (initial) process $\theta:\Omega\rightarrow C$ which is $\sig_0$-measurable. Let $L,l>0$, and consider a stock whose price at time $t$ is given by a process $\left(S^l(t)\right)_{t\in[0,T]}$ satisfying the stochastic functional differential equation (SFDE):
\begin{equation}\label{sfdel}
\left\{ \begin{array}{ll}
dS^l(t)=f(t,S^l_{t-l})S^l(t)dt+g(t,S^l_{t-l})S^l(t)dW(t), \quad t\in [0,T]\\
S^l(t)= \hat{\theta}(t), \quad t\in [-l-L,0],
\end{array}\right.
\end{equation}
where $\hat{\theta}$ is given by
\begin{equation*}\label{hattheta2}
\hat{\theta}(t):=\left\{ \begin{array}{ll}
\theta(t), \quad t\in [-L,0],\\
\theta(-L), \quad t\in [-l-L,-L].
\end{array}\right..
\end{equation*}
The process $W$ is a 1-dimensional Brownian Motion on $(\Omega,\sig,(\sig_t)_{t\in[0,T]},P)$ and, for any $t\in[-l,T]$, $S^l_t\in C$ is given by $S^l_t(s):=S(t+s)$, $s\in [-L,0]$. For $t\in[-1,0]$, define $\sig_t:=\sig_0.$

A solution of (\ref{sfdel}) is a sample continuous process $S^l:[-l-L,T]\times \Omega \rightarrow \real$ such that $S^l|_{[0,t]}$ is $(\sig_s)_{t\in[0,s]}$-adapted, $S^l(s)$ is $\sig_0$-measurable for all $s\in[-l-L,0]$, and $S^l$ satisfies the It\^{o} integral equation
\begin{equation*}
S^l(t)=\left\{ \begin{array}{ll}
\theta(\cdot)(0)+\int_0^t{f(u,S^l_{u-l})S(u)du}+\int_0^t{g(u,S_{u-l})S(u)dW(u)}, \quad t\in [0,T]\\
\hat{\theta}(\cdot)(t), \quad t\in [-l-L,0].
\end{array}\right.
\end{equation*}

\begin{remark} 
	The SFDE (\ref{sfdel}) is not a particular case of the existence theorem introduced in \cite{mohammed}. Moreover, as we will show in the proof of theorem \ref{e&ul}, the functionals $f$ and $g$ need only satisfy a joint continuity condition in order for the SFDE (\ref{sfdel}) to admit a global solution. This is an interesting gain in contrast with the continuity, local Lipschitz and global linear growth conditions imposed on the functionals in \cite{mohammed}. However, we shall see in section \ref{ndp} that in order to obtain convergence (as $l\rightarrow 0$) of solutions of (\ref{sfdel}) to a process with full finite memory, we must impose additional conditions on $f$ and $g$.
\end{remark}

\subsection{Existence and uniqueness of a feasible solution}
The next result provides the existence of a unique solution for the SFDE (\ref{sfdel}). Moreover, if $\theta(0)$ is strictly positive a.s., then so is the solution of $(\ref{sfdel})$. This is a very important feature, since $S^l$ describes a stock price.
\begin{theorem}\label{e&ul}
	Consider the framework of section \ref{framework2}. Then the SFDE (\ref{sfdel}) has a unique solution satisfying
	\begin{eqnarray}\label{soll}
	\nonumber S^l(t)&=&\theta(0)\exp{\left\{ \int_0^t{f(u,S^l_{u-l})du}+\int_0^t{g(u,S^l_{u-l})dW(u)}\right.}\\&-&\left. \frac{1}{2}\int_0^t{g(u,S^l_{u-l})^2du} \right\},\quad t\in [0,T].
	\end{eqnarray}
\end{theorem}

\noindent \emph{Proof.} We show this by induction in steps of length $l$. For simplicity, consider $T$ a multiple of $l$. For $t\in [0,l]$, we have
\begin{equation}\label{dsl}
dS^l(t)=S^l(t)[f(t,\hat{\theta}_{t-l})dt+g(t,\hat{\theta}_{t-l})dW(t)].
\end{equation}
Define the process
\[
N_1(t):=\int_0^t{f(u,\hat{\theta}_{u-l})du}+\int_0^t{g(u,\hat{\theta}_{u-l})dW(u)},\quad t\in [0,l].
\]
The continuity of $f$, $g$ and $\theta$ imply that the processes $f(t,\hat{\theta}_{t-l})=f(t,\cdot)\circ\hat{\theta}_{t-l}$ and $g(t,\hat{\theta}_{t-l})=g(t,\cdot)\circ\hat{\theta}_{t-l}$, $t\in[0,l]$, are $\sig_0$-measurable and continuous. Hence, the process
$A_1(t):=\int_0^t{f(u,\hat{\theta}_{u-l})du}$, $t\in[0,l]$, has almost all sample paths continuously differentiable. Also, from the sample path continuity of $g(t,\hat{\theta}_{t-l})$, $t\in[0,l]$, we have that 
\[
\int_0^t g(u,\hat{\theta}_{u-l})^2du<\infty\quad a.s.\quad \forall\, t\in[0,l].
\]

Hence, the process $M_1(t):=\int_0^t g(u,\hat{\theta}_{u-l})dW(u)$, $t\in[0,l]$, is a continuous $(\sig_t)_{t\in[0,l]}$-local martingale and therefore, the process $N_1(t):=A_1(t)+M_1(t)$, $t\in[0,l]$, is a continuous semimartingale.
Equation (\ref{dsl}) can then be written as a linear stochastic differential equation 
(SDE)
\begin{equation*}
\left\{ \begin{array}{ll}
dS^l(t)=S^l(t)dN_1(t), \quad t\in [0,l],\\
S^l(0)= \theta(0),
\end{array}\right.
\end{equation*}
which has a unique solution (Dol\'{e}ans-Dade exponential) given by
\begin{eqnarray*}
	S^l(t)&=&\theta(0)\exp\left\{N_1(t)-\frac{1}{2}[N_1,N_1](t)\right\}\\
	&=&\theta(0)\exp{\left\{ \int_0^t{f(u,\hat{\theta}_{u-l})du}+\int_0^t{g(u,\hat{\theta}_{u-l})dW(u)}\right.}\\&-&\left. \frac{1}{2}\int_0^t{g(u,\hat{\theta}_{u-l})^2du} \right\},\quad t\in [0,l].
\end{eqnarray*}
This implies that equation (\ref{soll}) holds for $t\in[0,l]$.

Now assume that equation (\ref{soll}) holds for $t\in [0,nl]$, where $n$ is a positive integer less than $T/l$. Then from equation (\ref{soll}), $\left(S^l(t)\right)_{t\in[0,nl]}$ is $(\sig_t)_{t\in[0,nl]}$-adapted and continuous. This implies that $(S^l_t)_{t\in[0,nl]}$ is also $(\sig_t)_{t\in[0,nl]}$-adapted and continuous (Lemma II-2.1 in Mohammed \cite{mohammed}). In other words, $S^l_{t-l}$ is continuous and $\sig_{t-l}$-measurable for $t\in[0,(n+1)l]$. 
Then the continuity of $f$ and $g$ imply that the processes $f(t,S^l_{t-l})=f(t,\cdot)\circ S^l_{t-l}$ and $g(t,S^l_{t-l})=g(t,\cdot)\circ S^l_{t-l}$, $t\in[0,(n+1)l]$, are $(\sig_{t})_{t\in[0,(n+1)l]}$-adapted and continuous. Hence, the process
$A_{n+1}(t):=\int_0^t{f(u,S^l_{u-l})du}$, $t\in[0,(n+1)l]$, has almost all sample paths continuously differentiable and 
$M_{n+1}(t):=\int_0^t g(u,S^l_{u-l})dW(u)$, $t\in[0,(n+1)l]$, is a continuous $(\sig_t)_{t\in[0,(n+1)l]}$-local martingale. Therefore, the process $N_{n+1}(t)=A_{n+1}(t)+M_{n+1}(t)$, $t\in[0,(n+1)l]$, is a semimartingale
and  the linear SDE
\begin{equation*}
\left\{ \begin{array}{ll}
dS^l(t)=S^l(t)dN_{n+1}(t), \quad t\in [0,(n+1)l]\\
S^l(0)= \theta(0),
\end{array}\right.
\end{equation*}
has a unique solution (Dol\'{e}ans-Dade exponential) given by
\begin{eqnarray*}
	S^l(t)
	&=&\theta(0)\exp{\left\{ \int_0^t{f(u,S^l_{u-l})du}+\int_0^t{g(u,S^l_{u-l})dW(u)}\right.}\\&-&\left. \frac{1}{2}\int_0^t{g(u,S^l_{u-l})^2du} \right\},\quad t\in [0,(n+1)l].
\end{eqnarray*}
Notice that if $\theta(0)>0$, then so is $S^l(t)$ for any $t\in[0,T]$. This completes the induction argument, and therefore equation (\ref{soll}) has a unique solution for any $t\in[0,T].\quad_\square$

\begin{theorem}\label{e&ul2}
	Consider the filtered probability space $(\Omega,\sig,(\sig_t)_{t\in[o,T]},P)$ and the 1-dimensional Brownian Motion $W$ on $(\Omega,\sig,(\sig_t)_{t\in[o,T]},P)$. Let 
	$\tilde{\theta}\in L^2(\Omega,C)$, $\tilde{f}:[0,T]\times L^2(\Omega,C)\rightarrow \real$ and $\tilde{g}:[0,T]\times L^2(\Omega,C)\rightarrow \real$, with $\tilde{f}$ and $\tilde{g}$ being jointly continuous and satisfying the linear growth condition:
	$$
	|\tilde{f}(t,\eta)|+|\tilde{g}(t,\eta)|\leq D(1+\|\eta\|_{L^2(\Omega,C)}),\quad \mbox{for any } t\in[0,T] \mbox{and } \eta\in C.
	$$
	The constant $D$ is independent of $t$ and $\eta$. Then the SFDE 
	\begin{equation*}\label{sfdelt}
	\left\{ \begin{array}{ll}
	dS^l(t)=\tilde{f}(t,S^l_{t-l})S^l(t)dt+\tilde{g}(t,S^l_{t-l})S^l(t)dW(t), \quad t\in [0,T]\\
	S^l(t)= \hat{\tilde{\theta}}(t), \quad t\in [-l-L,0],
	\end{array}\right.
	\end{equation*}
	has a unique nonnegative solution satisfying
	\begin{eqnarray*}
		\nonumber S^l(t)&=&\tilde{\theta}(0)\exp{\left\{ \int_0^t{\tilde{f}(u,S^l_{u-l})du}+\int_0^t{\tilde{g}(u,S^l_{u-l})dW(u)}\right.}\\&-&\left. \frac{1}{2}\int_0^t{\tilde{g}(u,S^l_{u-l})^2du} \right\},\quad t\in [0,T].
	\end{eqnarray*}
	The process $\hat{\tilde{\theta}}$ is given by
	\begin{equation*}
	\hat{\tilde{\theta}}(t):=\left\{ \begin{array}{ll}
	\tilde{\theta}(t), \quad t\in [-L,0],\\
	\tilde{\theta}(-L), \quad t\in [-l-L,-L].
	\end{array}\right..
	\end{equation*}
\end{theorem}

\noindent \emph{Proof.} The proof is similar to that of theorem \ref{e&ul}, but in addition, one needs to show that for any $t\in[-l,T]$, each $S^l_{t}$ is in $L^2(\Omega,C)$. We show this in proposition \ref{step1fm}, which uses 
a martingale-type inequality for the Ito integral, 
stated below. 

\begin{lemma}\label{mohisometry}
	Let $W:[a,b]\times\Omega\rightarrow\real^m$ be an $m$-dimensional Brownian Motion on a filtered probability space $(\Omega, \mathcal{F}, (\mathcal{F}_t)_{t\in[a,b]}, P)$. Suppose $g:[a,b]\times\Omega\rightarrow L(\real^m,\real^d)$ is measurable, $(\mathcal{F}_t)_{t\in[a,b]}$-adapted and $\int_a^b E|g(t,\cdot)|^{2k}dt<\infty$, for a positive integer $k\geq 1$. Then
	$$ E\sup_{t\in[a,b]}\left|\int_a^t g(u,\cdot)dW(u)\right|^{2k}\leq A_k(b-a)^{k-1}\int_a^b E|g(u,\cdot)|^{2k}du,$$
	where $$A_k:=d^{k-1}\left(\frac{4k^3m^2}{2k-1}\right)^k.$$
\end{lemma}
\noindent \emph{Proof.} For a proof, the reader may refer to Mohammed \cite{mohammed} (pg. 27).\\


\begin{proposition}\label{step1fm}
	Let $\tilde{\theta}$, $\tilde{f}$ and $\tilde{g}$ satisfy the assumptions of theorem \ref{e&ul2}. Then for each $t\in[0,T]$, the process $S^l$ satisfies
	\begin{equation*}
	E\left[\sup_{v\in[0,t]}|S^l(v)|^2\right]+\|S^l_{t-l}\|^2_{L^2(\Omega,C)}\leq U_{T/l},
	\end{equation*}
	where $U_{T/l}$ is a constant satisfying
	\begin{equation*}
	U_{T/l}\geq\left(E|\tilde{\theta}(0)|^{2}\right)^{T/l}(DT)^{2T/l}.
	\end{equation*}
	This shows that 
	\begin{equation*}
	E\left[\sup_{v\in[0,t]}|S^l(v)|^2\right]+\|S^l_{t-l}\|^2_{L^2(\Omega,C)}
	\end{equation*}
	is not uniformly bounded in $l$.
\end{proposition}
\noindent \emph{Proof.} For simplicity, consider $T$ a multiple of $l$. We use induction with steps of length $l$. More specifically, we show that for any $t\in[0,nl]$, $n=1,2,\dots,T/l$, 
\begin{equation*}
E\left[\sup_{v\in[0,t]}|S^l(v)|^2\right]+\|S^l_{t-l}\|^2_{L^2(\Omega,C)}\leq U_{n},
\end{equation*}
where $U_n$ is a constant satisfying
\begin{equation}\label{constun}
U_{n}\geq
\left(E|\tilde{\theta}(0)|^{2}\right)^{n}(DT)^{2n}.
\end{equation}
We first show that the proposition holds for any $t\in[0,l]$. Applying (in order) Jensen's inequality (finite and integral forms), lemma \ref{mohisometry}, and the linear growth property of $\tilde{f}$ and $\tilde{g}$ we have, for $t\in[0,l]$,
\begin{eqnarray}\label{calc1fm}
& & E\left [ \sup_{v\in[0,t]}|S^l(\cdot)(v)|^2\right ]\nonumber \\
&=& E\left [  \sup_{v\in[0,t]}\left| \tilde{\theta}(0)+\int_0^v{\tilde{f}(u,\hat{\tilde{\theta}}_{u-l})S^l(u)du}+\int_0^v{\tilde{g}(u,\hat{\tilde{\theta}}_{u-l})S^l(u)dW(u)} \right|^2 \right ]\nonumber \\
&\leq& E\left [  \sup_{v\in[0,t]}\left( 3\left| \tilde{\theta}(0) \right|^2+3\left| \int_0^v{\tilde{f}(u,\hat{\tilde{\theta}}_{u-l})S^l(u)du} \right|^2 \right. \right.\nonumber\\
& &+\,\, 3 \left. \left. \left|\int_0^v{\tilde{g}(u,\hat{\tilde{\theta}}_{u-l})S^l(u)dW(u)}\right|^2 \right) \right ] \nonumber \\
&\leq& E\left [ 3|\tilde{\theta}(0)|^2\right]+3tE\left [ \int_0^{t}{|\tilde{f}(u,\hat{\tilde{\theta}}_{u-l})|^2|S^l(u)|^2du}\right]\nonumber \\
& &+\,\, 3\cdot 4 E\left[\int_0^{t}|{\tilde{g}(u,\hat{\tilde{\theta}}_{u-l})|^2|S^l(u)|^2du}\right]\nonumber\\
&\leq& 3E|\tilde{\theta}(0)|^2+3D^2\left(t+4\right)E\int_0^{t}(1+\|\hat{\tilde{\theta}}_{u-l}\|_{L^2(\Omega,C)})^2|S^l(u)|^2du\nonumber\\
&\leq& 3E|\tilde{\theta}(0)|^2+3D^2(T+4)(1+\|\tilde{\theta}\|_{L^2(\Omega,C)})^2\int_0^tE\left [ \sup_{v\in[0,u]}|S^l(v)|^2\right ]du.\nonumber\\
\end{eqnarray}
Hence, from Gronwall's inequality, we obtain for $t\in[0,l]$:
\begin{eqnarray*}
	E\left [ \sup_{v\in[0,t]}|S^l(v)|^2\right ]\leq 3E|\tilde{\theta}(0)|^2 e^{3D^2(T+4)\left(1+\|\tilde{\theta}\|_{L^2(\Omega,C)}\right)^{2}t}.
\end{eqnarray*}
Also, for $t\in[0,l],$ $\|S_{t-l}^l\|_{L^2(\Omega,C)}=\|\hat{\tilde{\theta}}_{t-l}\|_{L^2(\Omega,C)}=\|\tilde{\theta}\|_{L^2(\Omega,C)}$, from which we obtain that for any $t\in[0,l]$, 
\begin{eqnarray*}
	& &E\left[\sup_{v\in[0,t]}|S^l(v)|^2\right]+\|S^l_{t-l}\|^2_{L^2(\Omega,C)}\\
	&\leq& \|\tilde{\theta}\|_{L^2(\Omega,C)}^2+3E|\tilde{\theta}(0)|^2 e^{3D^2(T+4)\left(1+\|\tilde{\theta}\|_{L^2(\Omega,C)}\right)^{2}T}=:U_1.
\end{eqnarray*} 
Notice that $U_1\geq E|\tilde{\theta}(0)|^2(DT)^2$. Now assume that proposition \ref{step1fm} holds for $t\in[0,nl]$, where $n$ is a positive integer $n<lT$. In particular, assume that for any $t\in[0,nl]$, 
$$E\left[\sup_{v\in[0,t]}|S^l(v)|^2\right]+\|S^l_{t-l}\|^2_{L^2(\Omega,C)}\leq U_n,$$
where $U_n$ is a positive constant satisfying (\ref{constun}).

Then for $t\in [l,(n+1)l]$,
\begin{eqnarray*}
	& & E \left[ \sup_{s\in [-L,0]}|S^l_{t-l}(s)|^2 \right] \leq E \left[ \sup_{s\in [-L-l,nl]}|S^l(s)|^2 \right]\\
	&\leq& E \left[ \sup_{s\in [-L,0]}|\tilde{\theta}(s)|^2 \right]+E \left[ \sup_{s\in [0,nl]}|S^l(s)|^2\right]\leq \|\tilde{\theta}\|^2_{L^2(\Omega,C)}+U_n.
\end{eqnarray*}
Further, in a calculation similar to (\ref{calc1fm}), we have that for any $t\in [l,(n+1)l]$,
\begin{eqnarray*}\label{calc2fm}
	& &E\left [ \sup_{v\in[0,t]}|S^l(\cdot)(v)|^2\right ]\nonumber \\ &\leq&3E|\tilde{\theta}(0)|^2+3D^2(T+4)(1+\|S^l_{u-l}\|_{L^2(\Omega,C)})^2\int_0^t|S^l(u)|^2du,\nonumber\\
	&\leq& 3E|\tilde{\theta}(0)|^2+3D^2(T+4)(1+\sqrt{U_n})^2 \int_0^tE\left[ \sup_{v\in[0,u]}|S^l(v)|^2\right]du.
\end{eqnarray*}
Hence, from Gronwall's inequality, we obtain
\begin{eqnarray*}
	E\left [ \sup_{v\in[0,t]}|S^l(v)|^2\right ]\leq 3E|\tilde{\theta}(0)|^2 e^{3D^2(T+4)(1+\sqrt{U_n})^{2}t }
\end{eqnarray*}
Thus, it follows that for any $t\in[0,(n+1)l]$, 
\begin{eqnarray*}
	& &E \left[\sup_{v\in[0,t]}|S^l(v)|^2\right] + \|S_{t-l}^l\|_{L^2(\Omega,C)}^2\\
	&\leq& \|\tilde{\theta}\|_{L^2(\Omega,C)}^2+U_n +3E|\tilde{\theta}(0)|^2 e^{3D^2(T+4)(1+\sqrt{U_n})^{2}T }=:U_{n+1}.
\end{eqnarray*}
Notice that 
\begin{eqnarray*}
	U_{n+1}&\geq& E|\tilde{\theta}(0)|^2(DT)^2U_n\geq E|\tilde{\theta}(0)|^2(DT)^2 (E|\tilde{\theta}(0)|^2)^n(DT)^{2n}\\
	&=&(E|\tilde{\theta}(0)|^2)^{n+1}(DT)^{2(n+1)}.
\end{eqnarray*}
Hence, proposition \ref{step1fm} holds for any $t\in [0,(n+1)l]$. This concludes the induction argument, with
\begin{equation*}
E \left[\sup_{v\in[0,t]}|S^l(v)|^2\right] + \|S^l_{t-l}\|_{L^2(\Omega,C)}^2\leq U_{T/l},\,\,\,t\in[0,T]
\end{equation*}
where $U_{T/l}$ is a constant satisfying $U_{T/l}\geq(E|\tilde{\theta}(0)|^2)^{T/l}(DT)^{2T/l}$. $\quad_\square$\\

In order to better understand the difference between the coefficients in theorems \ref{e&ul} and \ref{e&ul2}, consider the following examples.

\begin{example}\label{example1}
	Let 
	$f_1:[0,T]\times C\rightarrow \real\quad\mbox{and}\quad g_1:[0,T]\times C\rightarrow \real$
	be functionals that have linear growth and satisfy the conditions in section \ref{framework2}. Then the functionals 
	$f_2:[0,T]\times L^2(\Omega,C)\rightarrow \real\mbox{ and } g_2:[0,T]\times L^2(\Omega,C)\rightarrow \real$
	defined by
	\begin{eqnarray*}
		\quad f_2(t,\eta)&:=&\int_\Omega f_1(t,\eta(\omega))dP(\omega)=E f_1(t,\eta(\cdot)),\\
		\quad g_2(t,\eta)&:=&\int_\Omega g_1(t,\eta(\omega))dP(\omega)=E g_1(t,\eta(\cdot)),
	\end{eqnarray*}
	for any $t\in[0,T]$, $\eta\in C$, satisfy the hypothesis of theorem \ref{e&ul2}.
\end{example}

\begin{example}\label{exmple2}
	In this example, we assign a uniform mean to the drift term and a uniform standard deviation to the diffusion term. Let
	$f:[0,T]\times C\rightarrow \real$ and $g:[0,T]\times C\rightarrow \real$
	be defined by
	\begin{eqnarray*}
		f(t,\eta)&:=&\frac{1}{L}\int_{-L}^0 \eta(u)du, \quad t\in[0,T],\\
		g(t,\eta)&:=&\left(\frac{1}{L}\int_{-L}^0 (\eta(u)-f(u,\eta))^2du\right)^{1/2},
		\quad t\in[0,T],
	\end{eqnarray*}
	which are jointly continuous functionals satisfying the conditions in  theorem \ref{e&ul}.
\end{example}

\section{A stock price model with full finite memory}\label{ndp}

In this section we introduce a feasible stock price model in which the volatility and drift terms depend on a finite history of the stock price up to the present time. The existence of such model is obtained by ``closing the memory gap", i.e., by proving
convergence in $L^2(\Omega,C([-L,T],\real))$, as $l\rightarrow 0$, of the solution of the SFDE (\ref{sfdel}) to a process with full finite memory.

\subsection{Framework}\label{framework3}
Let $L>0$ and consider a stock whose price at time $t$ is given by a process $\left(S(t)\right)_{t\in[0,T]}$ satisfying the stochastic functional differential equation:
\begin{equation}\label{sfdesm}
\left\{ \begin{array}{ll}
dS(t)=f(t,S_t)S(t)dt+g(t,S_t)S(t)dW(t), \quad t\in [0,T]\\
S(t)= \theta(t), \quad t\in [-L,0],
\end{array}\right.
\end{equation}
on a filtered probability space $(\Omega,\sig,(\sig_t)_{t\in[0,T]},P)$ satisfying the usual conditions.  The initial process $\theta\in L^2(\Omega,C)$ is $\sig_0$-measurable.
The process $W$ is a 1-dimensional Brownian Motion on $(\Omega,\sig,(\sig_t)_{t\in[0,T]},P)$, and $S_t$ is given by $S_t(s):=S(t+s)$, $s\in [-L,0]$, for any $t\in [0,T]$. The functionals $f:[0,T]\times L^2(\Omega,C)\rightarrow \real$ and $g:[0,T]\times L^2(\Omega,C)\rightarrow \real$ are jointly continuous,  globally bounded and uniformly Lipschitz in the second variable, viz.
\begin{eqnarray}
&&\nonumber |f(t,\psi)|\leq f_{max}\quad \mbox{and}\quad |g(t,\psi)|\leq g_{max}\quad \mbox{ and } \\ 
&&|f(t,\psi_1)-f(t,\psi_2)|+|g(t,\psi_1)-g(t,\psi_2)|\leq \alpha \|\psi_1-\psi_2\|_{L^2(\Omega,C)}\label{lipschitzsm}
\end{eqnarray}
for all $t\in[0,T]$ and $\psi$, $\psi_1, \psi_2 \in L^2(\Omega,C)$. The Lipschitz constant $\alpha$ is independent of $t\in[0,T]$. 


A solution of (\ref{sfdesm}) is an $(\sig_t)_{t\in[0,T]}$-adapted process $S\in L^2(\Omega,C([-L,T],\real))$ starting off at $\theta$, 
and satisfying the It\^{o} integral equation
\begin{equation}\label{soldef}
S(t)=\left\{ \begin{array}{ll}
\theta(\cdot)(0)+\int_0^t{f(u,S_{u})S(u)du}+\int_0^t{g(u,S_{u})S(u)dW(u)}, \quad t\in [0,T]\\
\theta(\cdot)(t), \quad t\in [-L,0].
\end{array}\right.
\end{equation}


\begin{remark} 
	Even with the boundedness conditions imposed on $f$ and $g$, the SFDE (\ref{sfdesm}) does not satisfy the general existence and uniqueness conditions  introduced in \cite{mohammed}, i.e., the 
	functionals 
	$\tilde{f}(t,\eta):=f(t,\eta)\eta(0)$ and $\tilde{g}(t,\eta):=g(t,\eta)\eta(0)$, $t\in[0,T]$, $\eta\in L^2(\Omega,C)$, are not globally Lipschitz.
	It is possible to set a modified Lipschitz condition on $f$ and $g$ in order for the functionals $\tilde{f}$ and $\tilde{g}$ to be globally Lipschitz (see Chang \cite{chang}). However, in order to derive the option pricing formulas, we use the 
	model with memory gap and its convergence to a model with full finite memory. Therefore, we will give an existence and uniqueness proof by \emph{closing the memory gap} in the SFDE (\ref{sfdel}). In doing this, the  approximation scheme will be feasible stock price models. 
	
\end{remark}


\subsection{Existence and uniqueness of a feasible solution}
\begin{theorem}\label{e&u}
	Consider the framework of section \ref{framework3}. The SFDE (\ref{sfdesm}) has a unique solution satisfying
	\begin{eqnarray}\label{solsm}
	\nonumber S(t)&=&\theta(0)\exp{\left\{ \int_0^t{f(u,S_{u})du}+\int_0^t{g(u,S_{u})dW(u)}\right.}\\&-&\left. \frac{1}{2}\int_0^t{g(u,S_{u})^2du} \right\},\quad t\in [0,T].
	\end{eqnarray}
\end{theorem}
In order to prove theorem \ref{e&u}, we define the sequence of processes
\begin{equation}\label{sfdeksm}
S^k(t)=\left\{ \begin{array}{l}
\theta(0)+\int_0^t{f(u,S_{u-1/k}^k)S^k(u)du}\\
+\int_0^t{g(u,S_{u-1/k}^k)S^k(u)dW(u)},\quad   \mbox{if } t\in [0,T]\\
\hat{\theta}(t), \quad \mbox{if } t\in [-1-L,0],
\end{array}\right.
\end{equation}
$k\geq1.$ For $t\in[-1,T]$, the memory segment $S^k_t$ is given by $S^k_t(s):=S^k(t+s)$, $s\in[-L,0].$ For $t\in[-1,T]$, define $\sig_t:=\sig_0$. The process $\hat{\theta}$ is given by 
\begin{equation*}
\hat{\theta}(t)=\left\{ \begin{array}{ll}
\theta(t), \quad t\in [-L,0],\\
\theta(-L), \quad t\in [-1-L,-L].
\end{array}\right.
\end{equation*}
From section \ref{dp}, each $S^k$ exists uniquely and satisfies 
\begin{eqnarray*}\label{solsmk}
	\nonumber S^k(t)&=&\theta(0)\exp{\left\{ \int_0^t{f(u,S_{u-1/k}^k)du}+\int_0^t{g(u,S_{u-1/k}^k)dW(u)}\right.}\\&-&\left. \frac{1}{2}\int_0^t{g(u,S_{u-1/k}^k)^2du} \right\},\quad t\in [0,T].
\end{eqnarray*}
The proof of theorem \ref{e&u} follows from Propositions \ref{step1sm} through \ref{step6sm}.

\begin{proposition}\label{step1sm}
	For any $\gamma\geq 1$ and each $t\in[0,T]$, the process $S^k$ satisfies
	\begin{equation*}
	E\left[\sup_{v\in[0,t]}|S^k(v)|^{2\gamma}\right]\leq U_\gamma,
	\end{equation*}
	where $U_\gamma$ is a constant independent of $k$.
\end{proposition}
\noindent \emph{Proof.} For simplicity, consider  a positive integer $T$. Applying Jensen's inequality (finite and integral forms) and lemma \ref{mohisometry}, we have for any $t\in[0,T]$,
{\allowdisplaybreaks
	\begin{eqnarray}\label{calc1sm}
	& & E\left [ \sup_{v\in[0,t]}|S^k(\cdot)(v)|^{2\gamma}\right ]\nonumber \\
	&=& E  \sup_{v\in[0,t]}\left| \theta(0)+\int_0^v{f(u,S^k_{u-1/k})S^k(u)du}
	+\int_0^v{g(u,S^k_{u-1/k})S^k(u)dW(u)} \right|^{2\gamma}\nonumber \\
	&\leq& E  \sup_{v\in[0,t]}\left( 3^{2\gamma-1}\left| \theta(0) \right|^{2\gamma}+3^{2\gamma-1}\left| \int_0^v{f(u,S^k_{u-1/k})S^k(u)du} \right|^{2\gamma}\right.\nonumber\\
	& &+\,\,\left.3^{2\gamma-1} \left|\int_0^v{g(u,S^k_{u-1/k})S^k(u)dW(u)}\right|^{2\gamma} \right)  \nonumber \\
	&\leq& E\left [ 3^{2\gamma-1}|\theta(0)|^{2\gamma}\right]+3^{2\gamma-1}t^{2\gamma-1}E\left [ \int_0^{t}{|f(u,S^k_{u-1/k})|^{2\gamma}|S^k(u)|^{2\gamma}du}\right]\nonumber \\
	& &+\,\,3^{2\gamma-1}A_{\gamma}t^{\gamma-1}E\left[\int_0^{t}|{g(u,S^k_{u-1/k})|^{2\gamma}|S^k(u)|^{2\gamma}du}\right]\nonumber\\
	&\leq& 3^{2\gamma-1}E|\theta(0)|^{2\gamma}+3^{2\gamma-1}t^{\gamma-1}f_{max}^{2\gamma}\int_0^{t}E\left [ \sup_{v\in[0,u]}|S^k(v)|^{2\gamma}\right ]du\nonumber\\
	& &+\,\, 3^{2\gamma-1}A_{\gamma}t^{\gamma-1}g_{max}^{2\gamma}\int_0^{t}E\left [ \sup_{v\in[0,u]}|S^k(v)|^{2\gamma}\right ]du\nonumber\\
	&\leq& 3^{2\gamma-1}E|\theta(0)|^{2\gamma}+3^{2\gamma-1}T^{\gamma-1}(T^{\gamma}f_{max}^{2\gamma}+A_{\gamma}g_{max}^{2\gamma})\int_0^{t}E  \sup_{v\in[0,u]}|S^k(v)|^{2\gamma}du,\nonumber
	\end{eqnarray}
}
where $A_\gamma:=(\frac{4\gamma^3}{2\gamma-1})^\gamma$.
Hence, from Gronwall's inequality, we obtain for $t\in[0,T]$:
\begin{eqnarray*}
	E\left [ \sup_{v\in[0,t]}|S^k(v)|^{2\gamma}\right ]\leq 3^{2\gamma-1}E|\theta(0)|^{2\gamma} e^{3^{2\gamma-1}T^{\gamma-1}(T^{\gamma}f_{max}^{2\gamma}+A_{\gamma}g_{max}^{2\gamma})T}:=U_{\gamma}.\quad_\square
\end{eqnarray*}

\begin{remark}
	Proposition \ref{step1sm} gives a uniform bound on $E\left[\sup_{v\in[0,t]}|S^k(v)|^{2\gamma}\right]$. We were able to obtain such a bound from the global boundedness of $f$ and $g$. This result may be compared to the (non-uniform) bound obtained in proposition \ref{step1fm}, where only a linear growth condition was assumed for the drift and diffusion terms. 
\end{remark}


\begin{proposition}\label{step2sm}
	For any integer $\gamma\geq 1$, each $S^k$ satisfies $E|S^k(t)-S^k(s)|^{2\gamma}\leq B_\gamma|t-s|^\gamma$ for all $s,t\in[0,T]$, where $B_\gamma$ is a constant independent of $k$.
\end{proposition}
\noindent \emph{Proof.} By Jensen's inequality (finite and integral forms), lemma \ref{mohisometry} and proposition \ref{step1sm}, we obtain for any $0\leq s,t \leq T$:
\allowdisplaybreaks{
	\begin{eqnarray*}
		& & E\left[\left|S^k(t)-S^k(s)\right|^{2\gamma}\right]\\
		&=& E\left[\left|\int_s^tf(u,S^k_{u-1/k})S^k(u)du+\int_s^tg(u,S^k_{u-1/k})S^k(u)dW(u)\right|^{2\gamma}\right]\\
		&\leq& 2^{2\gamma-1}|t-s|^{2\gamma-1}E\int_s^t|f(u,S^k_{u-1/k})|^{2\gamma}|S^k(u)|^{2\gamma}du\\
		&+&2^{2\gamma-1}A_\gamma|t-s|^{\gamma-1}E\int_s^t|g(u,S^k_{u-1/k})|^{2\gamma}|S^k(u)|^{2\gamma}du\\
		&\leq&2^{2\gamma-1}(T^\gamma f_{max}^{2\gamma}+A_\gamma g_{max}^{2\gamma})|t-s|^{\gamma-1}\int_s^t E|S^k(u)|^{2\gamma}du\\
		&\leq& 2^{2\gamma-1}(T^\gamma f_{max}^{2\gamma}+A_\gamma g_{max}^{2\gamma})|t-s|^{\gamma-1}U_{\gamma}|t-s|=B_\gamma|t-s|^{\gamma},
	\end{eqnarray*}
	where $B_\gamma:=2^{2\gamma-1}(T^\gamma f_{max}^{2\gamma}+A_\gamma g_{max}^{2\gamma})U_{\gamma}$. 
	$\quad_\square$
}\\

Next, we state Kolmogorov's continuity criterion for a sequence of Banach-valued stochastic processes. The theorem will be used in 
proposition \ref{step3sm}.

\begin{theorem}\label{kolm}
	\textbf{(Kolmogorov's continuity criterion for a sequence of stochastic processes)}. Let $\{X^k(t)\}_{k=1}^\infty$, $t\in[0,T]$, be a sequence of stochastic processes with values in a Banach space $E$. Assume that there exist positive constants $\rho_1$, $c$ and $\rho_2>1$, all independent of $k$, satisfying
	\begin{equation*}\label{unifbound}
	E[\|X^k(t)-X^k(s)\|_E^{\rho_1}]\leq c|t-s|^{\rho_2},
	\end{equation*}
	for every $s,t\in [0,T]$. Then each $X^k$ has a continuous modification $\tilde{X}^k$.
	Further, let $b$ be an arbitrary positive number less than $\frac{\rho_2-1}{\rho_1}$. Then there exists a positive random variable $\xi_k$ with $E[\xi_k^{\rho_1}]<H$, where H is a constant independent of $k$, such that
	\[
	\|\tilde{X}^k(t)-\tilde{X}^k(s)\|_E\leq \xi_k|t-s|^b,
	\]
	for every $s,t\in [0,T]$ and a.s..
\end{theorem}
\noindent \emph{Proof.} The reader may refer to Kunita \cite{kunita}, pg. 31, for a proof. 

\begin{proposition}\label{step3sm}
	Let $\beta\in(0,1/2)$ be a fixed constant. Each $S^k$ satisfies
	\begin{eqnarray*}
		&(i)&\,\,|S^k(t)-S^k(s)|\leq c_k|t-s|^\beta\,\,\mbox{ for all } s,t\in[0,T]\,\, a.s.;\\
		&(ii)&\,\,\|S^k_t-S^k_s\|^2_{L^2(\Omega,C)}\leq 3\tilde{c}|t-s|^{2\beta}\\
		& & +\, 2E\sup_{v\in(-(t\wedge L)\wedge0,-(s\wedge L)\wedge0]}|\hat{\theta}(0)-\hat{\theta}(s+v)|^2\\
		& & +\,E\sup_{v\in[-L,-(t\wedge L)\wedge0]}|\hat{\theta}(t+v)-\hat{\theta}(s+v)|^2\,\,\mbox{ for all } -1\leq s<t\leq T,\,\, a.s.,
	\end{eqnarray*}
	where $\tilde{c}$ is a constant independent of $k$ and $c_k$ is a positive random variable satisfying $E(c_k^{2\rho})\leq \tilde{c}$ with $\rho$ being the smallest integer greater than $\frac{1}{1-2\beta}$.
\end{proposition}
\noindent \emph{Proof.}
Let $\rho$ be the smallest integer greater than $\frac{1}{1-2\beta}$. From proposition \ref{step2sm}, $E|S^k(t)-S^k(s)|^{2\rho}\leq B_\rho|t-s|^{\rho}$, for any $s,t\in[0,T]$. Since $\beta<\frac{\rho-1}{2\rho}$, then it follows from Kolmogorov's continuity criterion (theorem \ref{kolm}) that there exists a positive random variable $c_k$ such that $|S^k(t)-S^k(s)|\leq c_k|t-s|^\beta$ a.s. for all $s,t\in[0,T]$, with $E(c_k^{2\rho})\leq \tilde{c}$, where $\tilde{c}$ is a constant independent of $k$. This proves part (i).

We now proceed to prove part (ii). For any $-1\leq s<t\leq T$, 
\begin{eqnarray}
\nonumber \|S^k_t-S^k_s\|_{L^2(\Omega,C)}^2 &=& E\sup_{v\in[-L,0]}|S^k(t+v)-S^k(s+v)|^2\\ \label{es1sm}
&\leq&E\sup_{v\in(-(s\wedge L)\wedge0,0]}|S^k(t+v)-S^k(s+v)|^2\\ \label{es2sm}
&+&E\sup_{v\in(-(t\wedge L)\wedge0,-(s\wedge L)\wedge0]}|S^k(t+v)-S^k(s+v)|^2\hspace*{7mm}\\ \label{es3sm}
&+&E\sup_{v\in[-L,-(t\wedge L)\wedge0]}|S^k(t+v)-S^k(s+v)|^2.
\end{eqnarray}
Using part (i), the term  (\ref{es1sm}) becomes:
\begin{eqnarray}
\nonumber& &E\sup_{v\in(-(s\wedge L)\wedge0,0]}|S^k(t+v)-S^k(s+v)|^2\leq E \sup_{v\in(-(s\wedge L)\wedge0,0]}\left(c_k|t-s|^\beta\right)^2\\
\nonumber &=&E(c_k^2|t-s|^{2\beta})=E(c_k^2)|t-s|^{2\beta}\leq \tilde{c}|t-s|^{2\beta}.
\end{eqnarray}
Moreover, from part (i), we also obtain for (\ref{es2sm}):
\begin{eqnarray}
\nonumber& &E\sup_{v\in(-(t\wedge L)\wedge0,-(s\wedge L)\wedge0]}|S^k(t+v)-S^k(s+v)|^2\\
\nonumber&\leq& 2E\sup_{v\in(-(t\wedge L)\wedge0,-(s\wedge L)\wedge0]}\{|S^k(t+v)-S^k(0)|^2+|\hat{\theta}(0)-\hat{\theta}(s+v)|^2\}\\
\nonumber&\leq& 2E\sup_{v\in(-(t\wedge L)\wedge0,-(s\wedge L)\wedge0]}\{c_k^2|t+v|^{2\beta}+|\hat{\theta}(0)-\hat{\theta}(s+v)|^2\}\\
\nonumber&\leq& 2E(c_k^2)|t-s|^{2\beta}+2E\sup_{v\in(-(t\wedge L)\wedge0,-(s\wedge L)\wedge0]}|\theta(0)-\theta(s+v)|^2\\
\nonumber&\leq& 2\tilde{c}|t-s|^{2\beta}+2E\sup_{v\in(-(t\wedge L)\wedge0,-(s\wedge L)\wedge0]}|\hat{\theta}(0)-\hat{\theta}(s+v)|^2.
\end{eqnarray}
Finally, (\ref{es3sm}) becomes:
\begin{equation}
\nonumber E\sup_{v\in[-L,-(t\wedge L)\wedge0]}|S^k(t+v)-S^k(s+v)|^2=E\sup_{v\in[-L,-(t\wedge L)\wedge0]}|\hat{\theta}(t+v)-\hat{\theta}(s+v)|^2.
\end{equation}
Hence, for $-1\leq s<t\leq T,$
\newpage
\begin{eqnarray*}
	\|S^k_t-S^k_s\|_{L^2(\Omega,C)}^2 &\leq& \tilde{c}|t-s|^{2\beta}+2\tilde{c}|t-s|^{2\beta}\\
	&+&2E\sup_{v\in(-(t\wedge L)\wedge0,-(s\wedge L)\wedge0]}|\hat{\theta}(0)-\hat{\theta}(s+v)|^2\\
	&+&E\sup_{v\in[-L,-(t\wedge L)\wedge0]}|\hat{\theta}(t+v)-\hat{\theta}(s+v)|^2\\
	&=&3\tilde{c}|t-s|^{2\beta}+2E\sup_{v\in(-(t\wedge L)\wedge0,-(s\wedge L)\wedge0]}|\hat{\theta}(0)-\hat{\theta}(s+v)|^2\\
	&+&E\sup_{v\in[-L,-(t\wedge L)\wedge0]}|\hat{\theta}(t+v)-\hat{\theta}(s+v)|^2. \quad_\square
\end{eqnarray*}

\begin{proposition}\label{step4sm}
	The sequence $(S^k)_{k=1}^\infty$ converges to a limit $S\in L^2(\Omega,C([-1-L,T],\real))$. 
\end{proposition}
\noindent\emph{Proof}. We first notice that for any $t\in[-1,T]$,
\begin{eqnarray}\label{trajlksm}
\nonumber \|S^l_t-S^k_t\|^2_{L^2(\Omega,C)}&=&E\sup_{s\in[-L,0]}|S^l(t+s)-S^k(t+s)|^2\\
&\leq& E\sup_{s\in[-1-L,t]}|S^l(s)-S^k(s)|^2\nonumber\\
&=&E\sup_{s\in[0,t]}|S^l(s)-S^k(s)|^2.\quad \quad
\end{eqnarray}
Then, for any $t\in[0,T]$ and $l>k$, we have that
{\allowdisplaybreaks
	\begin{eqnarray}\label{convsm}
	\nonumber &E& \sup_{v\in[0,t]}|S^l(v)-S^k(v)|^2\\
	\nonumber &=& E\sup_{v\in[0,t]}\left|\int_0^v\{f(u,S^l_{u-1/l})S^l(u)-f(u,S^k_{u-1/k})S^k(u)\}du\right.\\
	\nonumber& &+\left.\int_0^v\{g(u,S^l_{u-1/l})S^l(u)-g(u,S^k_{u-1/k})S^k(u)\}dW(u)\right|^2\\
	\nonumber&\leq&2tE\int_0^t|f(u,S^l_{u-1/l})S^l(u)-f(u,S^k_{u-1/k})S^k(u)|^2du\\
	\nonumber& &+2\cdot 4 E\int_0^t|g(u,S^l_{u-1/l})S^l(u)-g(u,S^k_{u-1/k})S^k(u)|^2du\\
	\nonumber&\leq&2tE\int_0^t(2|f(u,S^l_{u-1/l})-f(u,S^k_{u-1/k})|^2|S^l(u)|^2\\
	\nonumber& &+2|f(u,S^k_{u-1/k})|^2|S^l(u)-S^k(u)|^2)du\\
	\nonumber& &+8E\int_0^t(2|g(u,S^l_{u-1/l})-g(u,S^k_{u-1/k})|^2|S^l(u)|^2\\
	\nonumber& &+2|g(u,S^k_{u-1/k})|^2|S^l(u)-S^k(u)|^2)du\\
	\nonumber&\leq&4\alpha^2(t+4)E\int_0^t\|S^l_{u-1/l}-S^k_{u-1/k}\|_{L^2(\Omega,C)}^2|S^l(u)|^2du\\
	\nonumber& &+4E\int_0^t(tf_{max}^2+4g_{max}^2)|S^l(u)-S^k(u)|^2du\\
	\nonumber&\leq& 8\alpha^2(t+4)\int_0^t(\|S^l_{u-1/l}-S^l_{u-1/k}\|_{L^2(\Omega,C)}^2\\
	\nonumber&&+\|S^l_{u-1/k}-S^k_{u-1/k}\|_{L^2(\Omega,C)}^2)E|S^l(u)|^2du\\
	\nonumber& &+4(tf_{max}^2+4g_{max}^2)^2\int_0^tE|S^l(u)-S^k(u)|^2du\\
	\nonumber&\leq&8\alpha^2U_1(t+4)\int_0^t(\|S^l_{u-1/l}-S^l_{u-1/k}\|_{L^2(\Omega,C)}^2+E\sup_{v\in[0,u]}|S^l(v)-S^k(v)|^2)du\\
	\nonumber& &+4(tf_{max}^2+4g_{max}^2)\int_0^t\sup_{v\in[0,u]}E|S^l(u)-S^k(u)|^2du\\
	\nonumber&\leq&4[2\alpha^2U_1(T+4)+Tf_{max}^2+4g_{max}^2]\int_0^tE\sup_{v\in[0,u]}|S^l(v)-S^k(v)|^2)du\\
	& &+8\alpha^2U_1(t+4)\int_0^t\|S^l_{u-1/l}-S^l_{u-1/k}\|_{L^2(\Omega,C)}^2du.
	\end{eqnarray}
	Let $C_1:=8\alpha^2U_1(t+4)$ and $C_2:=4[2\alpha^2U_1(T+4)+Tf_{max}^2+4g_{max}^2]$. Then, applying Gronwall's inequality to the above inequality, we get
	\begin{eqnarray}\label{gronwall1sm}
	E\sup_{v\in[0,t]}|S^l(v)-S^k(v)|^2&\leq& C_1e^{C_2t}\int_0^t\|S^l_{u-1/l}-S^l_{u-1/k}\|_{L^2(\Omega,C)}^2du.
	\end{eqnarray}
}
We now show that $\int_0^t\|S^l_{u-1/l}-S^l_{u-1/k}\|^2_{L^2(\Omega,C)}du\rightarrow 0$ as $l,k\rightarrow\infty$. 
From proposition \ref{step3sm} (ii), it follows that for any $u\in[0,T]$,
\begin{eqnarray}\label{lalalasm}
\nonumber& &\|S^l_{u-1/l}-S^l_{u-1/k}\|^2_{L^2(\Omega,C_d)}\leq 3\tilde{c}|1/k-1/l|^{2\beta}\\
\nonumber &+&2E\sup_{v\in(-\{(u-1/l)\wedge L\}\wedge0,-\{(u-1/k)\wedge L\}\wedge0]}|\hat{\theta}(0)-\hat{\theta}(u-1/k+v)|^2\\
&+&E\sup_{v\in[-L,-\{(u-1/l)\wedge L\}\wedge0]}|\hat{\theta}(u-1/l+v)-\hat{\theta}(u-1/k+v)|^2.
\end{eqnarray}
Let $\epsilon>0$. By the uniform continuity of $\hat{\theta}$, there exists $0<\delta<\epsilon$ such that
\begin{eqnarray*}
	|s_1-s_2|<\delta\Rightarrow |\hat{\theta}(s_1)-\hat{\theta}(s_2)|<\max\left\{\left(\frac{\epsilon}{6\tilde{c}}\right)^{\frac{1}{2\beta}},\sqrt{\frac{\epsilon}{6}}\right\}.
\end{eqnarray*}
Then, for any $l>k>\frac{1}{\delta}$,it follows that $\frac{1}{l}<\frac{1}{k}<\delta$ and therefore, for any $v\in(-\{(u-1/l)\wedge L\}\wedge0,-\{(u-1/k)\wedge L\}\wedge0]$, we have that $|u-1/k+v|\leq\frac{1}{k}-\frac{1}{l}<\delta$. 
Hence, for any $l>k>\frac{1}{\delta}$, it follows from (\ref{lalalasm}) that
\begin{eqnarray*}
	\|S^l_{u-1/l}-S^l_{u-1/k}\|^2_{L^2(\Omega,C_d)}\leq 3\tilde{c}\left[\left(\frac{\epsilon}{6\tilde{c}}\right)^{\frac{1}{2\beta}}\right]^{2\beta}+2\left(\sqrt{\frac{\epsilon}{6}}\right)^2+\left(\sqrt{\frac{\epsilon}{6}}\right)^2=\epsilon,
\end{eqnarray*}
for any $u\in[0,T]$. 
Hence, for any $l>k>\frac{1}{\delta}$,
\begin{eqnarray*}
	\int_0^t\|S^l_{u-1/l}-S^l_{u-1/k}\|^2_{L^2(\Omega,C_d)}du\leq\int_{0}^{t}\epsilon du\leq T\epsilon.
\end{eqnarray*}
This shows that $\int_0^t\|S^l_{u-1/l}-S^l_{u-1/k}\|^2_{L^2(\Omega,C)}du\rightarrow 0$ as $l,k\rightarrow\infty$. Therefore, from inequality (\ref{gronwall1sm}), 
$$E \sup_{v\in[0,t]}|S^l(v)-S^k(v)|^2\rightarrow 0\quad\mbox{as}\quad l,k\rightarrow\infty.$$
Therefore, the sequence $(S^k)_{k=1}^\infty$ is a Cauchy sequence in $L^2(\Omega,C([-1-L,T],\real))$ and therefore convergent to a limit $S\in L^2(\Omega,C([-1-L,T],\real))$. From (\ref{trajlksm}), it also follows that for each $t\in[0,T]$, $(S^k_t)_{k=1}^{\infty}$ converges to $S_t$ in $L^2(\Omega,C).\quad_\square$

\begin{proposition}\label{step5sm}
	The process $S|_{[-L,T]}$ satisfies the SFDE (\ref{sfdesm}) and can be written as (\ref{solsm}).
\end{proposition}
\noindent \emph{Proof.} To show this, we take limits as $k\rightarrow\infty$ in both sides of (\ref{sfdeksm}). The left-hand side of (\ref{sfdeksm}) converges to $S$ in $L^2(\Omega,C([-1-L,T],\real))$. Furthermore, $S$ is $(\sig_t)_{t\in[0,T]}$-adapted, since each $S_k$ is. Moreover, in a calculation similar to (\ref{convsm}),
{\allowdisplaybreaks
	\begin{eqnarray}\label{inconvsm}
	\nonumber & &E\sup_{v\in[0,t]}\left|\int_0^v\{f(u,S_u)S(u)-f(u,S^k_{u-1/k})S^k(u)\}du\right.\\
	\nonumber&+&\left.\int_0^v\{g(u,S_u)S(u)-g(u,S^k_{u-1/k})S^k(u)\}dW(u)\right|^2\\
	\nonumber&\leq&4[2\alpha^2U_1(T+4)+Tf_{max}^2+4g_{max}^2]\int_0^tE\sup_{v\in[0,u]}|S(v)-S^k(v)|^2)du\\
	&+&8\alpha^2U_1(t+4)\int_0^t\|S_u-S_{u-1/k}\|_{L^2(\Omega,C)}^2du.
	\end{eqnarray}
}
From the continuity of $[0,T]\ni t\mapsto S_t$, it follows that
\begin{eqnarray*}
	& &\int_0^t\|S_{u}-S_{u-1/k}\|^2_{L^2(\Omega,C)}=\\
	& &\int_0^{1/k}\|S_{u}-\theta\|^2_{L^2(\Omega,C)}+\int_{1/k}^t\|S_{u}-S_{u-1/k}\|^2_{L^2(\Omega,C)}\rightarrow 0 \,\, \mathrm{as} \,\,k\rightarrow\infty.
\end{eqnarray*}
Also, as seen previously,
\begin{eqnarray*}
	E\sup_{v\in[0,u]}|S(v)-S^k(v)|^2 \rightarrow 0 \,\, \mathrm{as} \,\,k\rightarrow\infty.
\end{eqnarray*}

Hence, (\ref{inconvsm}) converges to 0 as $k\rightarrow\infty$.
This shows that, for any $t\in[0,T]$, the right-hand side of (\ref{sfdeksm}) converges to (\ref{soldef}) 
in $L^2(\Omega,C([-L,T],\real))$ as $k\rightarrow\infty$. Therefore, $S|_{[-L,T]}$ satisfies the SFDE (\ref{sfdesm})
and the process\\
$N(t):=\int_0^t f(u,S_u)du+\int_0^t g(u,S_u)dW(u)$, $t\in [0,T]$, is an $(\sig_t)_{t\in[0,T]}$-adapted continuous semimartingale.
We can then apply It\^{o}'s formula for semimartingales to 
\begin{equation*}
\left\{ \begin{array}{ll}
dS(t)=S(t)dN(t), \quad t\in [0,T]\\
S(0)= \theta(0),
\end{array}\right.
\end{equation*}
which gives
\begin{eqnarray*}
	S(t)&=&\theta(0)\exp{\left\{ \int_0^t{f(u,S_{u})du}+\int_0^t{g(u,S_{u})dW(u)}\right.}\\&-&\left. \frac{1}{2}\int_0^t{g(u,S_{u})^2du} \right\},\quad t\in [0,T].\quad_\square
\end{eqnarray*}

\begin{proposition}\label{step6sm}
	(Uniqueness) If $\tilde{S}$ is an $(\sig_t)_{t\in[0,T]}$-adapted process satisfying (\ref{sfdesm}), then $\tilde{S}=S|_{[-L,T]}$ a.s..
\end{proposition}
\noindent \emph{Proof.} In a calculation similar to (\ref{convsm}), we find for the difference
\begin{eqnarray}\label{uniq}
\nonumber & &\|\tilde{S}-S|_{[-L,T]}\|_{L^2(\Omega,C([-L,T],\real))}^2\leq E\sup_{v\in[0,T]}|\tilde{S}(v)-S(v)|^2\\
\nonumber &=&E\sup_{v\in[0,T]}\left|\int_0^v\{f(u,\tilde{S}_u)\tilde{S}(u)-f(u,S_{u})S(u)\}du\right.\\
\nonumber&+&\left.\int_0^v\{g(u,\tilde{S}_u)\tilde{S}(u)-g(u,S_{u})S(u)\}dW(u)\right|^2\\
\nonumber&\leq&4[\alpha^2U_1(T+4)+Tf_{max}^2+4g_{max}^2]\int_0^TE\sup_{v\in[0,u]}|\tilde{S}(v)-S(v)|^2)du.
\end{eqnarray}
Hence, from Gronwall's inequality, it follows that 
\[
E\sup_{v\in[0,T]}|\tilde{S}(v)-S(v)|^2=0\Rightarrow \tilde{S}=S|_{[-L,T]}\,\,\, a.s..\quad_\square
\]


\begin{theorem}\label{ratesm}
	Let $\beta\in(0,1/2)$ be a fixed constant. If the initial process $\theta$ satisfies 
	\begin{equation}\label{aholsm}
	E|\theta(t)-\theta(s)|^{2\gamma}\leq C_{\theta}|t-s|^\gamma,
	\end{equation}
	for any $\gamma>1$, where $C_\theta$ is a positive constant, then $\theta$ is pathwise $\beta$-H\"{o}lder continuous and 
	\begin{equation*}
	E \sup_{v\in[0,t]}|S(v)-S^k(v)|^2\leq c\left(\frac{1}{k}\right)^{2\beta},
	\end{equation*}
	where c is a constant independent of $k$.
\end{theorem}
\noindent \emph{Proof.} 
Let $\rho>\frac{1}{1-2\beta}$ be an integer. From (\ref{aholsm}) and the fact that $\beta<\frac{\rho-1}{2\rho}$, Kolmogorov's continuity criterion (Theorem \ref{kolm}) implies that there exists a positive random variable $c_\theta$ such that $|\theta(t)-\theta(s)|\leq c_\theta|t-s|^\beta$ a.s., with $E(c_\theta^\gamma)\leq \tilde{c_\theta}$, where $\tilde{c_\theta}$ is a positive constant. That is, $\theta$ is pathwise $\beta$-H\"{o}lder continuous.

Then notice that $\hat{\theta}$ is also pathwise $\beta$-H\"{o}lder continuous. Indeed, for $-1-L\leq s<t\leq 0$, 
\begin{eqnarray*}
	|\hat{\theta}(t)-\hat{\theta}(s)|=\left\{\begin{array}{lll}
		|\theta(t)-\theta(s)|\leq c_\theta|t-s|^{\beta}, \quad -L\leq s,t \leq 0\\
		|\theta(t)-\theta(-L)|\leq c_\theta|t+L|^{\beta}\leq c_\theta|t-s|^{\beta}, \quad s<-L,\,\,t>-L;\\
		|\theta(-L)-\theta(-L)|=0\leq c_\theta|t-s|^{\beta},\quad -1-L\leq s,t <-L.
	\end{array}
	\right.
\end{eqnarray*}
Then proposition \ref{step3sm} (ii) and the $\beta$-H\"{o}lder continuity of $\theta$ imply that 
\begin{eqnarray*}
	\|S^k_t-S^k_s\|_{L^2(\Omega,C)}^2&\leq&3\tilde{c}|t-s|^{2\beta}+2E\sup_{v\in[-(t\wedge L),-(s\wedge L)]}c_\theta^2|s+v|^2\\
	&+&E\sup_{v\in[-L,-(t\wedge L)]}c_\theta^2|t-s|^2\\
	&\leq&3\tilde{c}|t-s|^{2\beta}+2E(c_\theta^2)|t-s|^{2\beta}+E(c_\theta^2)|t-s|^{2\beta}\\
	&=&3(\tilde{c}+1)|t-s|^{2\beta},
\end{eqnarray*}
for any $-1\leq s<t \leq T$. Hence, it follows that 
\begin{eqnarray}
\nonumber\int_0^t\|S^l_{u-1/l}-S^l_{u-1/k}\|^2_{L^2(\Omega,C)}du\leq3T(\tilde{c}+\tilde{c_\theta})\left|\frac{1}{k}-\frac{1}{l}\right|^{2\beta}.
\end{eqnarray}
Therefore, from inequality (\ref{gronwall1sm}), we obtain
\begin{equation*}
E \sup_{v\in[0,t]}|S^l(v)-S^k(v)|^2\leq C_1e^{C_2t}3T(\tilde{c}+\tilde{c_\theta})\left|\frac{1}{k}-\frac{1}{l}\right|^{2\beta}.
\end{equation*}
Finally, letting $l\rightarrow\infty$ and $c:=3C_1Te^{C_2t}$, we obtain
\begin{equation*}
E \sup_{v\in[0,t]}|S(v)-S^k(v)|^2\leq c\left(\frac{1}{k}\right)^{2\beta}.\quad_\square
\end{equation*}

Propositions \ref{step1sm}-\ref{step6sm}
complete the proof of theorem \ref{e&u}. Theorem \ref{ratesm} gives the order of convergence for the approximation scheme (\ref{sfdeksm}), when the initial process $\theta$ is $\beta$-H\"{o}lder continuous with $\beta\in(0,1/2)$. The approximation scheme (\ref{sfdeksm}) can be used as a numerical method for (\ref{sfdesm}). Notice that, if $\theta(0)$ is strictly positive, then so is the solution in theorem \ref{e&u}.

\section{An option pricing formula with \emph{memory gap}}\label{opdm}
In this section, we present an option pricing formula for the stock dynamics introduced in section  \ref{dp}. Such formula and its derivation are an extension of the ``Delayed Black-Scholes Formula" introduced in \cite{bs1}. 


Let $L,l>0$, $T$ be a multiple of $l$, and consider a stock whose price  at time $t$ is given by a process $S$ satisfying the SFDE 
\begin{equation}\label{sfdelt2}
\left\{ \begin{array}{ll}
dS^l(t)={f}(t,S^l_{t-l})S^l(t)dt+{g}(t,S^l_{t-l})S^l(t)dW(t), \quad t\in [0,T]\\
S^l(t)= \hat{\theta}(t), \quad t\in [-l-L,0],
\end{array}\right.
\end{equation}
where ${f}:[0,T]\times L^2(\Omega,C)\rightarrow \real$,  ${g}:[0,T]\times L^2(\Omega,C)\rightarrow \real$ and $\hat{\theta}$ satisfy the conditions of theorem \ref{e&ul2} (cf. $\tilde{f}$, $\tilde{g}$, $\hat{\tilde{\theta}}$). Additionally, assume that $\hat{\theta}$ is strictly positive and
\begin{equation}\label{hyp}
g(t,\eta)>0\quad \mbox{whenever}\quad \eta\quad \mbox{is strictly positive.}
\end{equation}
Assume also that  $\mathcal{F}_t=\mathcal{F}_t^W$ for all $t\in[0,T]$, where $(\mathcal{F}_t^W)_{t\in[0,T]}$ is the filtration generated by the Brownian Motion $W$. 

Considering that there are no transaction costs and that one can buy and sell stocks and bonds continuously in time, we arrive at the following theorem.

\begin{theorem}\label{fairl}
	Let $\{B,S^l\}$ be a market such that for a fixed $r\geq0$, $B(t)=e^{rt},\,t\in[0,T],$ and such that $S^l$ is described by the SFDE (\ref{sfdelt2}).
	Consider a contingent claim $Z$ on this market (e.g., the payoff of an option written on $S^l$ with maturity $T$).
	Then the market is complete and the fair price $V^l$ of $Z$ is given by
	$$V^l(t)=e^{-r(T-t)}E_{Q^l}[Z|\mathcal{F}_t^{S^l}],\quad t\in[0,T],$$ 
	where $Q^l$ is defined by $dQ^l=\rho^l(T)dP$ with $\rho^l(T)$ given by \begin{equation*}\label{rhol}
	\rho^l(t):=\exp{\left\{-\int_{0}^{T}{\frac{\{f(t,S^l_{t-l})-r\}}{g(t,S^l_{t-l})}dW(u)}-\frac{1}{2}\int_{0}^{T}{\left(\frac{\{f(t,S^l_{t-l})-r\}}{g(t,S^l_{t-l})}\right)^2du}\right\}}.
	\end{equation*} 
	The hedging strategy $\pi^l=(\pi_{S^l}^l,\pi_B^l)$ is given by
	$$\pi_{S^l}^l(t)=\frac{e^{rt}h^l(t)}{{S}^l(t)g(t,S^l_{t-l})},\quad \pi_B^l(t)=e^{-rT}\{E_{Q^l}[Z|\mathcal{F}_t^{S^l}]-\pi_{S^l}^l(t){S}^l(t)\},\quad t\in[0,T],$$
	where $h^l$ is given by 
	\begin{equation*}\label{Ml}
	M^l(t)=E_{Q^l}[e^{-rT}Z]+\int_0^t{h^l(u)dW^l(u)},\quad t\in[0,T].
	\end{equation*}
\end{theorem}

\noindent\emph{Proof.} 
With the goal of applying Girsanov's theorem \cite{girsanov}, we define the process $$X^l(t):=-\frac{\{f(t,S^l_{t-l})-r\}}{g(t,S^l_{t-l})}, \quad t\in[0,T].$$
Since we are assuming that $\theta$ is strictly positive a.s., it follows from theorem \ref{e&ul} that $S^l_{t-l}$ is strictly positive for all $t\in[0,T]$ a.s.. Then, by (\ref{hyp}), $X^l$ is (a.s.) well defined. 
In the proof of theorem \ref{e&ul} we saw that the processes $[0,T]\ni t\mapsto f(t,S^l_{t-l})$ and $[0,T]\ni t\mapsto g(t,S^l_{t-l})$ are continuous and $(\mathcal{F}_{t-l})_{t\in[0,T]}$-adapted. 
Hence $X^l(t)$ is $\mathcal{F}_{t-l}$-measurable and $\int_0^t{|X^l(u)|^2du}<\infty$ a.s. for each $t\in[0,T]$.  This implies that the stochastic integral 
$$\int_{T-kl}^{T-(k-1)l}{X^l(u)dW(u)}, \quad k=1,2,\ldots,\frac{T}{l},$$ 
conditioned on $\mathcal{F}_{T-kl}$, have normal distribution with mean zero and variance $$\int_{T-kl}^{T-(k-1)l}{X^l(u)^2du}.$$(Engel \cite{engel}). Then, by the formula for the moment generating function of a normal distribution we have:
\begin{eqnarray*}
	E_P\left[\exp{\left\{\int_{T-kl}^{T-(k-1)l}{X^l(u)dW(u)}\right\}}\bigg|\mathcal{F}_{T-kl}\right]=\exp{\left\{\frac{1}{2}\int_{T-kl}^{T-(k-1)l}{X^l(u)^2du}\right\}},
\end{eqnarray*}
for $k=1,2,\ldots,\frac{T}{l}$. Hence
\begin{eqnarray}\label{E1}
E_P\left[\exp{\left\{\int_{T-kl}^{T-(k-1)l}{X^l(u)dW(u)}-\frac{1}{2}\int_{T-kl}^{T-(k-1)l}{X^l(u)^2du}\right\}}\bigg|\mathcal{F}_{T-kl}\right]=1,\,\,
\end{eqnarray}
for $k=1,2,\ldots,\frac{T}{l}$, where it was used that $\exp{\left\{-\frac{1}{2}\int_{T-kl}^{T-(k-1)l}{X^l(u)^2du}\right\}}$ is $\mathcal{F}_{T-kl}$-measurable.
Now define the random variables $$Z_k:=\exp{\left\{\int_{0}^{T-kl}{X^l(u)dW(u)}-\frac{1}{2}\int_{0}^{T-kl}{X^l(u)^2du}\right\}},\quad k=1,2,\ldots,\frac{T}{l},$$ and note that for each $k$, $Z_k$ is $\mathcal{F}_{T-kl}$-measurable. 
By inductively conditioning $\exp{\left\{\int_{0}^{T}{X^l(u)dW(u)}-\frac{1}{2}\int_{0}^{T}{X^l(u)^2du}\right\}}$ on $\mathcal{F}_{T-l}$, $\mathcal{F}_{T-2l}$, and so on, it follows from (\ref{E1}) that
\begin{eqnarray}\label{induction}
\nonumber E_P\left[\exp{\left\{\int_{0}^{T}{X^l(u)dW(u)}-\frac{1}{2}\int_{0}^{T}{X^l(u)^2du}\right\}}\bigg|\mathcal{F}_{T-kl}\right]=Z_k
\end{eqnarray}
a.s. for $k=1,2,\ldots,\frac{T}{l}$. In particular, 
\begin{eqnarray}\label{E0}
E_P\left[\exp{\left\{\int_{0}^{T}{X^l(u)dW(u)}-\frac{1}{2}\int_{0}^{T}{X^l(u)^2du}\right\}}\bigg|\mathcal{F}_{0}\right]=1.
\end{eqnarray}
Taking expectation on both sides of (\ref{E0}) we arrive at
\begin{equation}\label{egirl}
E_P\left[\exp{\left\{\int_{0}^{T}{X^l(u)dW(u)}-\frac{1}{2}\int_{0}^{T}{X^l(u)^2du}\right\}}\right]=1.
\end{equation}
Hence, Girsanov's theorem  applies to the process $X^l$, and therefore
$$W^l(t):=W(t)+\int_0^t{\frac{\{f(u,S^l_{u-l})-r\}}{g(u,S^l_{u-l})}du},\qquad t\in[0,T],$$
is a Wiener process under the measure $Q^l$ defined by $dQ^l:=\rho^l(T)dP$, where
\begin{equation*}\label{rhol}
\rho^l(t):=\exp{\left\{-\int_{0}^{T}{\frac{\{f(t,S^l_{t-l})-r\}}{g(t,S^l_{t-l})}dW(u)}-\frac{1}{2}\int_{0}^{T}{\left(\frac{\{f(t,S^l_{t-l})-r\}}{g(t,S^l_{t-l})}\right)^2du}\right\}}.
\end{equation*}
Now let 
\begin{equation}\label{stilde1}
\tilde{S}^l(t):=\frac{S^l(t)}{B(t)}=e^{-rt}S^l(t), \quad t\in[0,T]
\end{equation} 
be the discounted stock price process. Then, by It\^o's formula it follows that
{\allowdisplaybreaks
	\begin{eqnarray*}
		d\tilde{S}^l(t)
		&=&\tilde{S}^l(t)\left[(f(t,S^l_{t-l})-r)dt+g(t,S^l_{t-l})dW(t)\right], \quad t\in[0,T],
	\end{eqnarray*}}
	which can be written as 
	\begin{eqnarray}\label{tilde}
	d\tilde{S}^l(t)&=&\tilde{S}^l(t)d\hat{S}^l(t), \quad t\in[0,T],
	\end{eqnarray}
	where $\hat{S}^l(t)$ is a continuous semimartingale defined by $$\hat{S}^l(t):=\int_{0}^{t}\{{f(u,S^l_{u-l})-r\}du}+\int_{0}^{t}{g(u,S^l_{u-l})dW(u)}, \quad t\in[0,T].$$
	Therefore, the process $\hat{S}^l$ can be written in the form
	\begin{equation}\label{shat}
	\hat{S}^l(t)=\int_0^t{g(u,S^l_{u-l})dW^l(u)},\qquad t\in[0,T].
	\end{equation}
	This implies that $\hat{S}^l$ is a continuous local martingale on $(\Omega,\mathcal{F},(\mathcal{F}_t)_{t\geq0},Q^l)$, and by (\ref{tilde}), the discounted stock price $\tilde{S}^l$ is also a continuous local martingale on $(\Omega,\mathcal{F},(\mathcal{F}_t)_{t\geq0},Q^l)$. In other words, $Q^l$ is an equivalent local martingale measure for $\tilde{S}^l$. Thus ,
	the market $\{B,S^l\}$ satisfies the no arbitrage property.
	We now establish the completeness of the market $\{B,S^l\}$. 
	
	From equations (\ref{stilde1}), (\ref{tilde}) and (\ref{shat}),
	we can write $$d\tilde{S}^l(t)= \tilde{S}^l(t)g(t,S^l_{t-l})dW^l(t),\,\,\,t\in[0,T].$$
	Since $[0,T]\ni t\mapsto\frac{1}{\tilde{S}^l(t)}$ and $[0,T]\ni t\mapsto\frac{1}{g(t,S^l_{t-l})}$ are continuous and $(\sig_t)_{t\in[0,T]}$-adapted, it follows that $\mathcal{F}_t=\mathcal{F}_t^{W^l}=\mathcal{F}_t^{\tilde{S}^l}=\mathcal{F}_t^{S^l}$.
	
	Let $Z$ be a contingent claim 
	and consider the $Q^l$-martingale
	$$M^l(t):=E_{Q^l}[e^{-rT}Z|\mathcal{F}_t^{S^l}]=E_{Q^l}[e^{-rT}Z|\mathcal{F}_t^{W^l}],\quad t\in[0,T].$$
	By the martingale representation theorem (see, e.g. \cite{protter}),
	\begin{equation*}\label{Ml}
	M^l(t)=E_{Q^l}[e^{-rT}Z]+\int_0^t{h^l(u)dW^l(u)},\quad t\in[0,T],
	\end{equation*}
	where $h^l$ is an $(\mathcal{F}_t^{W^l})_{t\in[0,T]}$-predictable process satisfying
	$$\int_0^t{|h^l(u)|^2ds}<\infty\,\,\,\mbox{a.s.}$$
	Define
	$$\pi_{S^l}^l(t):=\frac{h^l(t)}{\tilde{S}^l(t)g(t,S^l_{t-l})},\quad \pi_B^l(t):=M^l(t)-\pi_{S^l}^l(t)\tilde{S}^l(t),\quad t\in[0,T].$$
	The pair $\pi^l:=(\pi_B^l,\pi_{S^l}^l)$ is a \emph{trading strategy }
	because\\
	(i) $\pi_B^l$ and $\pi_{S^l}^l$ are $\{\mathcal{F}_t^{S^l}\}_{t\in[0,T]}$-predictable\\
	(ii) the stochastic integral $\int_0^T{\pi_{S^l}^l(t)d\tilde{S}^l(t)}$ exists since
	$$\int_0^T{\pi_{S^l}^l(t)d\tilde{S}^l(t)}=\int_0^T{\frac{h^l(t)}{\tilde{S}^l(t)g(t,S^l_{t-l})}\tilde{S}^l(t)g(t,S^l_{t-l})dW^l(t)}=\int_0^t{h^l(t)dW^l(t)}.$$
	Its \emph{value process} 
	is given by
	\begin{eqnarray*}
		V_{\pi^l}^l(t)&=&\pi_B^l(t)e^{rt}+\pi_{S^l}^l(t)S^l(t)=[M^l(t)-\pi_{S^l}^l(t)\tilde{S}^l(t)]e^{rt}+\frac{h^l(t)}{\tilde{S}^l(t)g(t,S^l_{t-l})}S^l(t)\\
		&=&M^l(t)e^{rt}-\frac{h^l(t)}{\tilde{S}^l(t)g(t,S^l_{t-l})}e^{rt}\tilde{S}^l(t)+\frac{h^l(t)e^{rt}}{g(t,S^l_{t-l})}=M^l(t)e^{rt}.
	\end{eqnarray*}
	which implies that the discounted value process is given by $\tilde{V}_{\pi^l}(t)=M^l(t)\,\,\,$for all$\,\,t\in[0,T]$.
	Then we have
	\begin{eqnarray*}
		d\tilde{V}_{\pi^l}(t)=dM^l(t)=h^l(t)dW^l(t)=\pi_{S^l}^l(t)\tilde{S}^l(t)g(t,S^l_{t-l})dW^l(t)=\pi_{S^l}^l(t)d\tilde{S}^l(t),
	\end{eqnarray*}
	which means that $\{\pi_B^l,\pi_{S^l}^l\}$ is a \emph{self-financing strategy}. 
	Also, 
	$$V^l(T)=e^{rT}M^l(t)=e^{rT}E_{Q^l}[e^{-rT}Z|\mathcal{F}_T^{W^l}]=e^{rT}e^{-rT}Z=Z,$$
	since $Z$ is $\mathcal{F}_T^{W^l}$-measurable. Hence $Z$ is \emph{attainable} and therefore, the market is complete.
	This implies that in order for the augmented market $\{B,S^l,Z\}$ to satisfy the no arbitrage property, the price of the claim $Z$ at time $t\in[0,T]$ must be
	$$V^l(t)=e^{rt}E_{Q^l}[e^{-rT}Z|\mathcal{F}_t^{S^l}]=e^{-r(T-t)}E_{Q^l}[Z|\mathcal{F}_t^{S^l}].\quad_\square$$ 
	
	
	The next theorem is a particular case of theorem \ref{fairl}, when the claim $Z$ is the payoff of an European call option written on the stock $S$ with maturity time $T$.
	
	\begin{theorem}\label{opthl}
		Assume the market $\{B,S^l\}$ satisfies the conditions of theorem \ref{fairl} and let $V^l(t)$ be the fair price at time $t$ of a European call option written on the stock $S^l$ with exercise price $K$ and maturity time $T$. Let $\Phi$ denote the distribution function of a standard normal variable, i.e.,
		$$\Phi(x):=\frac{1}{\sqrt{2\pi}}\int_{-\infty}^x{e^{-u^2/2}du},\quad x\in \real.$$
		Then for all $t\in[T-l,T]$, $V^l(t)$ is given by
		\begin{equation}\label{cform}
		V^l(t)=S^l(t)\Phi(\beta_+(t))-Ke^{-r(T-t)}\Phi(\beta_-(t)),
		\end{equation}
		where
		$$\beta_{\pm}(t):=\frac{\ln\left(\frac{S^l(t)}{K}\right)+\int_t^T\left(r\pm\frac{1}{2}g(u,S^l_{u-l})^2\right)du}{\sqrt{\int_t^T{g(u,S^l_{u-l})^2du}}}.$$
		If $T>l$ and $t<T-l$, then
		$$V^l(t)=e^{rt}E_{Q^l}\left[H\left(\tilde{S}^l(T-l), -\frac{1}{2}\int_{T-l}^T{g(u,S^l_{u-l})^2du}, \int_{T-l}^T{g(u,S^l_{u-l})^2du}\right)\bigg|\mathcal{F}_t\right],$$
		where $H$ is given by
		\begin{eqnarray*}
			H(x,m,\sigma^2)&:=&xe^{m+\sigma^2/2}\Phi\left(\frac{\sigma^2+\ln\left(\frac{x}{K}\right)+rT+m}{\sigma}\right)\\
			& &-Ke^{-rT}\Phi\left(\frac{\ln\left(\frac{x}{K}\right)+rT+m}{\sigma}\right),
		\end{eqnarray*}
		for $\sigma,x\in \real^+, m\in \real.$
		The hedging strategy is given by
		$$\pi_{S^l}^l(t)=\Phi(\beta_+(t)),\quad \pi_B^l(t)=-Ke^{-rT}\Phi(\beta_-(t)),\quad t\in[T-l,T].$$
	\end{theorem}
	\emph{Proof.} 
	%
	Taking $Z=(S^l(t)-K)^+$ in theorem \ref{fairl} and using that $\mathcal{F}_t=\mathcal{F}_t^{S^l}$, the fair price of the option at time $t$ is given by
	$$V^l(t)=e^{-r(T-t)}E_{Q^l}[(S^l(t)-K)^+|\mathcal{F}_t]=e^{rt}E_{Q^l}[(\tilde{S}^l(T)-Ke^{-rT})^+|\mathcal{F}_t],\quad t\in[0,T].$$
	From theorem \ref{e&ul} we have
	\begin{eqnarray*}
		\tilde{S}^l(t)&=&e^{-rt}S^l(t)=\theta(0)\exp{\left\{ \int_0^t{\{f(u,S^l_{u-l})-r\}du}+\int_0^t{g(u,S^l_{u-l})dW(u)}\right.}\\
		& &\hspace*{4cm}-\left. \frac{1}{2}\int_0^t{g(u,S^l_{u-l})^2du} \right\}\\
		&=&\theta(0)\exp{\left\{ \int_0^t{g(u,S^l_{u-l})dW^l(u)}-\frac{1}{2}\int_0^t{g(u,S^l_{u-l})^2du}\right\}}, \quad t\in [0,T],
	\end{eqnarray*}
	which implies that
	$$\tilde{S}^l(T)=\tilde{S}^l(t)\exp{\left\{ \int_t^T{g(u,S^l_{u-l})dW^l(u)}-\frac{1}{2}\int_t^T{g(u,S^l_{u-l})^2du}\right\}}, \quad t\in [0,T].
	$$
	Hence
	\begin{eqnarray*}
		V^l(t)=e^{rt}E_{Q^l}\left[\left(\tilde{S}^l(t)\exp{\left\{ \int_t^T{g(u,S^l_{u-l})dW^l(u)}-\frac{1}{2}\int_t^T{g(u,S^l_{u-l})^2du}\right\}}\right.\right.\\
		-Ke^{-rT}\bigg)^+\bigg|\mathcal{F}_t\bigg],\quad t\in[0,T].
	\end{eqnarray*}
	If $t\in[T-l,T]$ then $\int_t^T{g(u,S^l_{u-l})^2du}$ is $\mathcal{F}_t$-measurable, so
	when conditioned on $\mathcal{F}_t$, the distribution of $\int_t^T{g(u,S^l_{u-l})dW^l(u)}$ under $Q^l$ is normal with mean zero and variance $\tilde{\sigma}^2:=\int_t^T{g(u,S^l_{u-l})^2du}.$ 
	That is, $\int_t^T{g(u,S^l_{u-l})dW^l(u)}$ has the same distribution as that of $\tilde{\sigma}^2\xi$, where $\xi\sim N(0,1)$. Since $\tilde{S}^l(t)$ is $\mathcal{F}_t$-measurable, we have
	\begin{eqnarray*}
		V^l(t)=e^{rt}E_{Q^l}\left[\left(\tilde{S}^l(t)e^{-\frac{1}{2}\tilde{\sigma}^2+\tilde{\sigma}\xi}-Ke^{-rT}\right)^+\right]=e^{rt}\tilde{H}\left(\tilde{S}^l(t),-\frac{1}{2}\tilde{\sigma}^2,\tilde{\sigma}^2\right),
	\end{eqnarray*}
	where
	$$\tilde{H}(x,m,\sigma^2):=E_{Q^l}[(xe^{m+\sigma\xi}-Ke^{-rT})^+],\quad \sigma, x \in \real^+,\, m\in \real.$$
	From a simple computation, it follows that $\tilde{H}=H$, as defined in the theorem. 
	Therefore, for $t\in[T-l,T],$
	{\allowdisplaybreaks
		\begin{eqnarray}\label{op}
		\nonumber V^l(t)&=&e^{rt}{H}\left(\tilde{S}^l(t),-\frac{1}{2}\tilde{\sigma}^2,\tilde{\sigma}^2\right)\\
		&=&S^l(t)\Phi(\beta_+(t))-Ke^{-r(T-t)}\Phi(\beta_-(t)).
		\end{eqnarray}}
	For $T>l$ and $t<T-l$ we have
	\begin{eqnarray*}
		V^l(t)&=&e^{rt}E_{Q^l}[(\tilde{S}^l(T)-Ke^{-rT})^+|\mathcal{F}_{t}]\\
		&=&e^{rt}E_{Q^l}\left[E_{Q^l}[(\tilde{S}^l(T)-Ke^{-rT})^+|\mathcal{F}_{T-l}]\big|\mathcal{F}_{t}\right]\\
		&=&e^{rt}E_{Q^l}\left[H\left(\tilde{S}^l(T-l), -\frac{1}{2}\int_{T-l}^T{g(u,S^l_{u-l})^2du}, \int_{T-l}^T{g(u,S^l_{u-l})^2du}\right)\bigg|\mathcal{F}_t\right].
	\end{eqnarray*}
	
	We now look for a closed form representation of the self-financing strategy $\pi^l=(\pi_B^l,\pi_{S^l}^l)$ in the last delay period. 
	Since $\pi^l$ is self-financing, we have
	\begin{eqnarray}\label{eq1}
	\nonumber dV^l_{\pi^l}(t)&=&\pi_B^l(t)dB(t)+\pi_{S^l}^l(t)dS^l(t)\\
	\nonumber&=&\pi_B^l(t)re^{rt}dt+\pi_{S^l}^l(t)\{f(t,S^l_{t-l})S^l(t)dt+g(t,S^l_{t-l})S^l(t)dW(t)\}\\
	\nonumber&=&\{\pi_B^l(t)re^{rt}+\pi_{S^l}^l(t)f(t,S^l_{t-l})S^l(t)\}dt\\
	& &+\pi_{S^l}^l(t)g(t,S^l_{t-l})S^l(t)dW(t).\hspace*{1.6cm}
	\end{eqnarray}
	The option price $V^l(t)$ at time $t\in[T-l,T]$ is given by (\ref{op}), which depends on $S(t)$, $S_{t-l}$ and $t$. Since $S_{t-l}$ is $\sig_{T-l}$-measurable for $t\in[T-l,T]$ (known), we can look at $V^l(t)$ as a function of $t$ and $S^l(t)$, viz. $V^l(t,S^l(t))$, where
	\begin{equation}\label{vx}
	V^l(t,x):=x\Phi(\beta_+(t,x))-Ke^{-r(T-t)}\Phi(\beta_-(t,x)), \quad t\in[T-l,T],\,\,x\in \real.
	\end{equation}
	From (\ref{op}), $V^l\in C^{1,2}([T-l,T]\times \real,\, \real)$, so we can apply the generalized It\^o's formula (pg. 92 in \cite{kunita}) to get
	\begin{eqnarray}\label{eq2}
	\nonumber && dV^l(t,S^l(t))
	=\left\{\frac{\partial{V^l}}{\partial{t}}(t,S^l(t))+\frac{\partial{V^l}}{\partial{x}}(t,S^l(t))f(t,S^l_{t-l})S^l(t)\right.\\
	& &+\left.\frac{1}{2}\frac{\partial^2{V^l}}{\partial{x^2}}(t,S^l(t))g(t,S^l_{t-l})^2S^l(t)^2\right\}dt+\frac{\partial{V^l}}{\partial{x}}(t,S^l(t))g(t,S^l_{t-l})S^l(t)dW(t).\hspace*{7.3cm}
	\end{eqnarray}
	Recall that
	we have $V^l_{\pi^l}(t)=V(t,S^l(t))$. Then, by  uniqueness of the representations (\ref{eq1}) and (\ref{eq2}) (Baxter \cite{baxter}), we must have that for $t\in[T-l,T]$,
	\begin{eqnarray}\label{pis}
	\left\{ \begin{array}{lll}
	\pi_{S^l}^l(t)g(t,S^l_{t-l})S^l(t)=\frac{\partial{V^l}}{\partial{x}}(t,S^l(t))g(t,S^l_{t-l})S^l(t),\\
	\pi_B^l(t)re^{rt}+\pi_{S^l}^l(t)f(t,S^l_{t-l})S^l(t)=\frac{\partial{V^l}}{\partial{t}}(t,S^l(t))\\
	\hspace*{1cm}+\frac{\partial{V^l}}{\partial{x}}(t,S^l(t))f(t,S^l_{t-l})S^l(t)
	+\frac{1}{2}\frac{\partial^2{V^l}}{\partial{x^2}}(t,S^l(t))g(t,S^l_{t-l})^2S^l(t)^2.
	\end{array}\right.
	\end{eqnarray}
	Hence, since $g(t,S^l_{t-l}),\,S^l(t)\,>0\,\,$for all $t\in[T-l,T]$, it follows that
	\begin{eqnarray*}\label{pisp}
		\pi_{S^l}^l(t)&=&\frac{\partial{V^l}}{\partial{x}}(t,S^l(t)), \quad t\in[T-l,T].
	\end{eqnarray*}
	With a simple calculation one can show that $$\frac{\partial{V^l}}{\partial{x}}(t,x)=\Phi(\beta_+(t,x)),\quad t\in[T-l,T].$$
	%
	This completes the proof.$\quad_\square$
	
	\begin{remark}
		Let $\,\,\mu,\sigma>0$. If $f(t,\eta)=\mu$ and $g(t,\eta)=\sigma$ in (\ref{cform}), for $t\in[0,T]$, $\eta\in C([-L,0],\real)$, we obtain the classical Black-Scholes model.
	\end{remark}

	\section{An option pricing formula with full (finite) memory}\label{opfm}
	In this section, we present the option pricing formula (as a conditional expectation) for the stock dynamics introduced in section  \ref{ndp}. The functionals $f$ and $g$ satisfy the assumptions in section  \ref{ndp} and, additionally,  
	there exists a constant $g_{min}$ such that 
	\begin{equation}\label{hyp2}
	g(t,\eta)>g_{min}\quad \mbox{whenever}\quad \eta \quad\mbox{is strictly positive.}
	\end{equation}
	Assume also that  $\mathcal{F}_t=\mathcal{F}_t^W$ for all $t\in[0,T]$, where $(\mathcal{F}_t^W)_{t\in[0,T]}$ is the filtration generated by the Brownian Motion $W$. 
	
	Considering that there are no transaction costs and that one can buy and sell stocks continuously on time, we arrive at the following theorem.
	
	\begin{theorem}\label{fair}
		Let $\{B,S\}$ be a market such that for fixed $r\geq0$, $B(t)=e^{rt},\,t\in[0,T],$ and such that $S$ is described by the SFDE (\ref{sfdesm}) with $\theta(t)>0\,\,\,$for all $\, t\in[-L,0]$ a.s..
		Consider a contingent claim $Z$ on this market.
		Then the market is complete and the fair price $V$ of $Z$ is given by
		$$V(t)=e^{-r(T-t)}E_Q[Z|\mathcal{F}_t^{S}],\quad t\in[0,T],$$ 
		where $Q$ is defined by $dQ=\rho(T)dP$ with $\rho(T)$ given by \begin{equation*}\label{rho}
		\rho(t):=\exp{\left\{-\int_{0}^{T}{\frac{\{f(t,S_{t})-r\}}{g(t,S_{t})}dW(u)}-\frac{1}{2}\int_{0}^{T}{\left(\frac{\{f(t,S_{t})-r\}}{g(t,S_{t})}\right)^2du}\right\}}.
		\end{equation*}
		The hedging strategy $\pi=(\pi_S,\pi_B)$ is given by
		$$\pi_S(t)=\frac{e^{rt}h(t)}{{S}(t)g(t,S_{t})},\quad \pi_B(t)=e^{-rT}\{E_Q[Z|\mathcal{F}_t^{S}]-\pi_S(t){S}(t)\},\quad t\in[0,T],$$
		where $h$ is given by 
		\begin{equation*}\label{M}
		M(t)=E_Q[e^{-rT}Z]+\int_0^t{h(u)d\hat{W}(u)},\quad t\in[0,T].
		\end{equation*}
		When  the claim $Z$ is the payoff of an European call option  written on the stock $S$ with exercise price $K$ and maturity time $T$, the fair price of the option is given by
		\begin{eqnarray}\label{opFull}
			V(t)=e^{rt}E_Q\left[\left(\tilde{S}(t)\exp{\left\{ \int_t^T{g(u,S_{u})d\hat{W}(u)}-\frac{1}{2}\int_t^T{g(u,S_{u})^2du}\right\}}\right.\right.\\
			-Ke^{-rT}\bigg)^+\bigg|\mathcal{F}_t\bigg],\quad t\in[0,T].
		\end{eqnarray}
	\end{theorem}
	
	\noindent\emph{Proof.}
	Define the processes
	$$X^k(t):=-\frac{\{f(t,S^k_{t-1/k})-r\}}{g(t,S^k_{t-1/k})}\quad\mbox{and}\quad X(t):=-\frac{\{f(t,S_{t})-r\}}{g(t,S_{t})},\quad t\in[0,T],$$
	for any positive integer $k$.
	Both processes are well defined since $S^k$ and $S$ are strictly positive. Moreover, $X^k$ and $X$ are continuous and $(\mathcal{F}_t)_{t\in[0,T]}$-adapted. From (\ref{egirl}), we have that
	\begin{equation}\label{egir1}
	E_P\left[\exp{\left\{\int_{0}^{T}{X^k(u)dW(u)}-\frac{1}{2}\int_{0}^{T}{X^k(u)^2du}\right\}}\right]=1.
	\end{equation}
	We also wish to apply Girsanov's theorem on the process $X$, i.e., we want to show that 
	$$E_P\left[\exp{\left\{\int_{0}^{T}{X(u)dW(u)}-\frac{1}{2}\int_{0}^{T}{X(u)^2du}\right\}}\right]=1.$$
	Let $h:[0,T]\times L^2(\Omega,C)\rightarrow \real$ be the functional defined by $h(t,\eta):=-\frac{f(t,\eta)-r}{g(t,\eta)}$ and notice that $h$ is uniformly bounded: 
	\begin{eqnarray*}
		|h(t,\eta))|&\leq& \frac{f_{max}+r}{g_{min}}.
	\end{eqnarray*}
	Define the sequence of processes $\{Y^k\}_{k=1}^{\infty}$ by
	\begin{equation}\label{sfdegirk}
	\left\{ \begin{array}{ll}
	dY^k(t)=h(t,S^k_{t-1/k})Y^k(t)dW(t), \quad t\in [0,T]\\
	Y^k(t)= 1, \quad t\in [-1-L,0],
	\end{array}\right.
	\end{equation}
	which has a unique nonnegative solution satisfying
	\begin{eqnarray}\label{solgirk}
	\nonumber Y^k(t)
	&=&\exp\left\{\int_0^t h(u,S^k_{u-1/k})dW(u)-\frac{1}{2}\int_0^t h(u,S^k_{u-1/k})^2du \right\}\\
	&=&\exp\left\{\int_0^t X^k(u)dW(u)-\frac{1}{2}\int_0^t X^k(u)^2du \right\},\quad t\in [0,T].
	\end{eqnarray}
	Now consider the process $\left(Y(t)\right)_{t\in[0,T]}$ defined by 
	\begin{equation}\label{sfdegir}
	\left\{ \begin{array}{ll}
	dY(t)=h(t,S_{t})Y(t)dW(t), \quad t\in [0,T]\\
	Y(t)= 1, \quad t\in [-L,0].
	\end{array}\right.
	\end{equation}
	
	The sequence $Y^k|_{[-L,0]}$ converges to $Y$ in $L^2(\Omega,C([-L,T],\real))$ and $Y(t)$ is given by
	\begin{eqnarray}\label{solgir}
	\nonumber Y(t)&=&\exp\left\{\int_0^t h(u,S_{u})dW(u)-\frac{1}{2}\int_0^t h(u,S_{u})^2du \right\}\\
	&=&\exp\left\{\int_0^t X(u)dW(u)-\frac{1}{2}\int_0^t X(u)^2du \right\},\quad t\in [0,T].
	\end{eqnarray}
	This is a particular case of the proof of theorem \ref{e&u}.
	
	As a consequence of such convergence and
	equation (\ref{egir1}), it follows  that
	\begin{eqnarray*}
		|EY(T)-1|^2&=&|EY(T)-EY^k(T)|^2\leq E|Y(T)-Y^k(T)|^2\\
		&\leq& E\sup_{t\in[0,T]}|Y(t)-Y^k(t)|^2\rightarrow 0\quad\mbox{as}\quad k\rightarrow\infty.
	\end{eqnarray*}
	This implies that $EY(T)=1$, or $$E_P\left[\exp{\left\{\int_{0}^{T}{X(u)dW(u)}-\frac{1}{2}\int_{0}^{T}{X(u)^2du}\right\}}\right]=1.$$
	Hence, Girsanov's theorem  applies to the process $X$, and therefore
	$$\hat{W}(t):=W(t)+\int_0^t{\frac{\{f(u,S_{u})-r\}}{g(u,S_{u})}du},\qquad t\in[0,T],$$
	is a Wiener process under the measure $Q$ defined by $dQ:=\rho(T)dP$, where
	\begin{equation*}\label{rho}
	\rho(t):=\exp{\left\{-\int_{0}^{T}{\frac{\{f(t,S_{t})-r\}}{g(t,S_{t})}dW(u)}-\frac{1}{2}\int_{0}^{T}{\left(\frac{\{f(t,S_{t})-r\}}{g(t,S_{t})}\right)^2du}\right\}}.
	\end{equation*}
	Let 
	\begin{equation}\label{stilde1f}
	\tilde{S}(t):=\frac{S(t)}{B(t)}=e^{-rt}S(t), \quad t\in[0,T],
	\end{equation} 
	be the discounted stock price process and define
	$$\hat{S}(t):=\int_{0}^{t}\{{f(u,S_{u})-r\}du}+\int_{0}^{t}{g(u,S_{u})dW(u)}, \quad t\in[0,T].$$
	Then, It\^o's formula implies that
	\begin{eqnarray}\label{tildef}
	\nonumber d\tilde{S}(t)
	&=&\tilde{S}(t)\left[(f(t,S_{t})-r)dt+g(t,S_{t})dW(t)\right]\\
	&=&\tilde{S}(t)d\hat{S}(t), \quad t\in[0,T].
	\end{eqnarray}
	Finally, note that the process $\hat{S}$ can be written in the form
	\begin{equation}\label{shatf}
	\hat{S}(t)=\int_0^t{g(u,S_{u})d\hat{W}(u)},\qquad t\in[0,T].
	\end{equation}
	Thus, $\hat{S}$ is a continuous local martingale on $(\Omega,\mathcal{F},(\mathcal{F}_t)_{t\geq0},Q)$, and by (\ref{tildef}), the discounted stock price $\tilde{S}$ is also a continuous local martingale on $(\Omega,\mathcal{F},(\mathcal{F}_t)_{t\geq0},Q)$. In other words, $Q$ is an equivalent local martingale measure for $\tilde{S}$. This implies 
	that the market \{B,S\} satisfies the no arbitrage property.
	
	Note that  $\mathcal{F}_t=\mathcal{F}_t^{\hat{W}}=\mathcal{F}_t^{\tilde{S}}=\mathcal{F}_t^{S}$.
	Now let $Z$ be a contingent claim  and consider the $Q$-martingale
	$$M(t):=E_Q[e^{-rT}Z|\mathcal{F}_t^{S}]=E_Q[e^{-rT}Z|\mathcal{F}_t^{\hat{W}}],\quad t\in[0,T].$$
	Then by the martingale representation theorem,
	\begin{equation*}\label{M}
	M(t)=E_Q[e^{-rT}Z]+\int_0^t{h(u)d\hat{W}(u)},\quad t\in[0,T],
	\end{equation*}
	where $h$ is an $(\mathcal{F}_t^{\hat{W}})_{t\in[0,T]}$-predictable process satisfying
	$$\int_0^t{|h(u)|^2ds}<\infty\,\,\,\mbox{a.s.}$$
	Define the trading strategy $\pi:=(\pi_B,\pi_S)$ by
	$$\pi_S(t):=\frac{h(t)}{\tilde{S}(t)g(t,S_{t})},\quad \pi_B(t):=M(t)-\pi_S(t)\tilde{S}(t),\quad t\in[0,T].$$
	Its value process  is given by
	\begin{eqnarray*}
		V_{\pi}(t)&=&\pi_B(t)e^{rt}+\pi_S(t)S(t)
		=M(t)e^{rt},
	\end{eqnarray*}
	which implies that the discounted value process is given by  $\tilde{V}_{\pi}(t)=M(t)\,\,\,$for all$\,\,t\in[0,T]$.
	Then we have
	\begin{eqnarray*}
		d\tilde{V}_{\pi}(t)=dM(t)=h(t)d\hat{W}(t)=\pi_S(t)\tilde{S}(t)g(t,S_{t})d\hat{W}(t)=\pi_S(t)d\tilde{S}(t),
	\end{eqnarray*}
	which means that $\{\pi_B,\pi_S\}$ is a self-financing strategy. Also, 
	$$V(T)=e^{rT}M(T)=e^{rT}E_Q[e^{-rT}Z|\mathcal{F}_T^{\hat{W}}]=e^{rT}e^{-rT}Z=Z$$
	since $Z$ is $\mathcal{F}_T^{\hat{W}}$-measurable. Hence $Z$ is attainable (and therefore the market is complete). Then 
	in order for the augmented market \{B,S,Z\} to satisfy the no arbitrage property, the price of the claim $Z$ at time $t\in[0,T]$ must be
	$$V(t)=e^{rt}E_Q[e^{-rT}Z|\mathcal{F}_t^{S}]=e^{-r(T-t)}E_Q[Z|\mathcal{F}_t^{S}].$$ 
	
	\noindent Taking $Z=(S(t)-K)^+$ and setting the stock price to the solution obtained in theorem \ref{e&u}, the fair price of the option at time $t$ is given by
		\begin{eqnarray*}
			V(t)=e^{rt}E_Q\left[\left(\tilde{S}(t)\exp{\left\{ \int_t^T{g(u,S_{u})d\hat{W}(u)}-\frac{1}{2}\int_t^T{g(u,S_{u})^2du}\right\}}\right.\right.\\
			-Ke^{-rT}\bigg)^+\bigg|\mathcal{F}_t\bigg],\quad t\in[0,T].\quad_\square
		\end{eqnarray*}

\begin{remark}
	The option pricing formula  (\ref{opFull}) allows one to use Monte Carlo methods to estimate the fair option price of a European option when the stock dynamics follows (\ref{sfdesm}). The stock dynamics (\ref{sfdesm}) can be seen as a generalized Geometric Brownian Motion with full finite memory.
\end{remark}

\bibliographystyle{amsplain}

\end{document}